\documentclass[usegraphicx,usedcolumn,usenatbib]{mn2e}
\usepackage{astbibref}
\usepackage{color,epsfig,amssymb,longtable}
\usepackage{array,colortbl}

\def\HII{{H{\sc ii}}}

\begin{document}

\title
{Kinematics of gas  and stars in the circumnuclear starforming ring of 
NGC\,3351}
\author[G. F. H\"agele et al.]
{Guillermo F. H\"agele$^{1}$\thanks{PhD fellow of Ministerio de Educaci\'on y
    Ciencia, Spain; guille.hagele@uam.es},
\'Angeles I. D\'{\i}az$^{1}$\thanks{on sabbatical leave at IoA, Cambridge},
 M\'onica V. Cardaci$^{1,2}$\thanks{PhD fellow of
  Ministerio de Educaci\'on y Ciencia, Spain}, 
  Elena Terlevich$^{3}$\thanks{Research
  Affiliate at IoA} \newauthor and Roberto Terlevich$^{3}$\thanks{Research
  Affiliate at IoA}
\\
$^{1}$ Departamento de F\'{\i}sica Te\'orica, C-XI, Universidad Aut\'onoma de
Madrid, 28049 Madrid, Spain\\ 
$^{2}$ XMM Science Operations Centre, European Space Astronomy Centre of ESA,
P.O. Box 50727, 28080 Madrid, Spain\\ 
$^{3}$ INAOE, Tonantzintla, Apdo. Postal 51, 72000 Puebla, M\'exico\\ }

\maketitle

\begin{abstract}
We have measured gas and stellar velocity dispersions in 5
circumnuclear starforming regions (CNSFRs) and the
nucleus of the barred spiral galaxy  NGC\,3351. The stellar dispersions have
been obtained from high resolution spectra of the CaT lines at
$\lambda\lambda$\,8494, 8542, 8662\,\AA , while the gas velocity dispersions
have been measured by Gaussian fits to the H$\beta$\,$\lambda$\,4861\,\AA\
line on high dispersion spectra.  

The CNSFRs, with sizes of about 100 to 150\,pc in diameter, are seen to be
composed of several individual star clusters with sizes between 1.7 and
4.9\,pc on an HST image. Using the stellar velocity dispersions, we have
derived dynamical masses for the entire starforming complexes and for the
individual star clusters. Values of the stellar velocity dispersions are
between 39 and 67\,km\,s$^{-1}$. Dynamical masses for the whole CNSFRs are
between 4.9\,$\times$\,10$^6$ and 4.3\,$\times$\,10$^7$\,M$_\odot$ and
between 1.8 and 8.7\,$\times$\,10$^6$\,M$_\odot$ for the individual star
clusters. 

Stellar and gas velocity dispersions are found to differ by about
20\,km\,s$^{-1}$ with the H$\beta$ lines being narrower than both the stellar
lines and the [O{\sc iii}]\,$\lambda$\,5007\,\AA\ lines. We have found
indications for the presence of two different kinematical components in the
ionised gas of the regions. The radial velocity curve shows deviation from
circular motions for the ionised hydrogen  consistent with its infall towards
the central regions of the galaxy at a velocity of about 25\,km\,s$^{-1}$. To
disentangle the origin of these two components it will be necessary to map
these regions with high spectral and spatial resolution and much better S/N in
particular for the O$^{2+}$ lines.

\end{abstract}


\section{Introduction}

The bulges of some nearby spiral galaxies show intense starforming regions
located in a  roughly annular pattern around their nuclei. At optical
wavelengths, these circumnuclear starforming regions (CNSFRs) are easily
observable rings. In the ultraviolet (UV), massive stars dominate the observed
circumnuclear emission even in the presence of an active nucleus
\citep{1998ApJ...505..174G,2002ApJ...579..545C}. \cite{2001ApJ...558...81C},
for a representative sample of 35 Seyfert 2 galaxies,  find that about 40\,\%
of them show unambiguous evidence of circumnuclear star formation within
300\,pc of the nucleus and that these starforming regions contribute about 30
to 50\,\% to the H$\beta$ total emission of the central zone.   
 
These CNSFRs, with sizes going from a few tens to a few hundreds of pc
\citep[e.g.][]{2000MNRAS.312..130D} seem to be made of several \HII\ regions
ionised by luminous compact stellar clusters whose sizes, as measured from
high spatial resolution HST images, are seen to be of only a few pc. Their
masses, as derived with the use of population synthesis models in
circumnuclear regions of different galaxies, suggest that these  clusters are
gravitationally bound and that they might evolve into globular cluster
configurations \citep{1996AJ....111.2248M}. The luminosities of CNSFRs are
rather large with absolute visual magnitudes (M$_v$) between -12 and -17 and
H$\alpha$ luminosities 
between 2\,$\times$\,10$^{38}$ and 7\,$\times$\,10$^{40}$ erg s$^{-1}$ . These
values are comparable to those shown by 30 Dor, the largest \HII\ region in the
LMC, and overlap with those shown by \HII\ galaxies 
\citep[][and references
  therein]{1988MNRAS.235..297M,2000MNRAS.311..120D,2006MNRAS.365..454H}. 

Although these \HII\ regions are very luminous not much is known about their
kinematics or dynamics for both the ionised gas and the stars. It could be
said that the worst known parameter of these ionising clusters is their mass. 
There are different methods to estimate the mass of a stellar
cluster. Classically one assumes that the system is virialized and determines
the total mass inside a radius by applying the virial theorem to the observed
velocity dispersion of the stars ($\sigma_{\ast}$). The stellar velocity
dispersion is however hard to measure in young stellar clusters (a few Myr old)
due to the 
shortage of prominent stellar absorption lines. The optical continuum between
3500 and 7000\,\AA\ shows very few lines since the light at  these wavelengths
is dominated by OB stars which have weak absorption lines at the same
wavelengths of the nebular emission lines (Balmer H and HeI lines). As
pointed out by several authors \citep[e.g.][]{1996ApJ...466L..83H}, at longer 
wavelengths (\,8500\,\AA) the contamination due to nebular lines is much
smaller  and since red supergiant stars, if present, dominate the near-IR
light where the Ca{\sc ii} $\lambda\lambda$\,8498, 8542, 8662\,\AA\ triplet
lines (CaT) are found, these should be easily observable allowing the
determination of $\sigma_{\ast}$
\citep{1990MNRAS.242P..48T,1994A&A...288..396P}. We have previously detected
the CaT lines in CNSFRs but at a spectral resolution that was below that
required to measure accurately their velocity dispersions
\citep[e.g.][]{1990MNRAS.242P..48T}. 

Added interest in the study of CNSFRs comes from the fact that they
are in general of high metal abundance (D\'\i az et al.~2006), 
therefore they provide clues for the 
understanding of star formation phenomena at large metallicities, and,
being close to the nuclear regions, for the
determination of metallicity gradients in spiral galaxies. 

NGC\,3351 is an SBb(r)II spiral galaxy \citep{1987rsac.book.....S}  with
coordinates  $\alpha_{\rm 2000}$=10$^h$\,43$^m$\,57\fs7, 
$\delta_{\rm 2000}$=+11$^{\circ}$\,42\arcmin\,12\farcs7. It was  classified as
a hot-spot galaxy by \cite{1967PASP...79..152S}. \cite{1982A&AS...50..491A}
studied the star formation activity in the nuclear region and  along a ring of
about 20\arcsec\ in diameter using broad and narrow band images, concluding
that NGC\,3351  harbours high-mass star formation in these zones. In fact it
can be considered a nuclear starburst galaxy since the star formation rate per
unit area in the nuclear region, as estimated from the H$\alpha$ emission,
compared to that in the disk is significantly increased
\citep{1992AJ....103..784D}. More recently, near infrared photometry in  the J
and K bands has been presented by \cite{1997AJ....114.1850E} who derive a
circumnuclear star formation rate of 0.38\,M$_\odot$\,yr$^{-1}$.   From CO
emission observations, Planesas et al.\ (1997) derive a mass of molecular gas
of 3.5\,$\times$\,10$^{8}$\,M$_\odot$.

HST UV images show that the present star formation in NGC\,3351 is
distributed along a nuclear ring with a  major axis of 15\arcsec\  with
starforming regions 
arranged in complexes of diameters between 1.4 and 2.0\arcsec\ which are made
up of several high surface brightness knots a few pc in size embedded in a
more diffuse component \citep{1997ApJ...484L..41C}. No signs of activity have
been observed in the nucleus of NGC\,3351. 

There are several values in the literature for the distance to NGC\,3351. Here
we adopt the distance derived by Graham et al.\ (1997) from Cepheid variable
stars as part of the Hubble Space Telescope (HST) Key Project on the
Extragalactic Distance Scale, which is  10.05\,Mpc giving a linear scale of
$\sim$\,49\,pc per arcsec. 

In this paper we present high-resolution  far red spectra
($\sim$0.39\,\AA\,px$^{-1}$\,$\sim$\,13.66\,km\,s$^{-1}$\,px$^{-1}$ at central
wavelength, $\lambda_c$=8563\,\AA ) and  stellar velocity dispersion
measurements ($\sigma_{\ast}$) along the line of sight, for the five
circumnuclear  starforming 
regions and the nucleus of the spiral galaxy NGC\,3351. We have also measured
the ionised gas velocity dispersions ($\sigma_{g}$) from high-resolution blue
spectra ($\sim$0.21\,\AA\,px$^{-1}$\,$\sim$\,12.63\,km\,s$^{-1}$\,px$^{-1}$ at
$\lambda_c$\,=\,4989\,\AA) using Balmer H$\beta$ and [O{\sc iii}] emission
lines. The comparison between $\sigma_{\ast}$ and $\sigma_{g}$  might throw
some light on the yet unsolved issue about the validity of the gravitational
hypothesis for the origin of the supersonic motions observed in the ionised
gas in Giant \HII\ regions \citep*{1999MNRAS.302..677M}. 

Section 2 presents the details of the observations and the data
reduction. Section 3 presents the results concerning the  kinematics of gas and
stars in each of the observed regions, as well as the determination of their
masses. Section 4 is devoted to the discussion of these results and finally
the summary and conclusions of this work are presented in Section 5. 

\nocite{1997ApJ...477..535G} 
\nocite{1997A&A...325...81P}

\section{Observations and data reduction}
\label{Obs}

\subsection{Observations}
  

\begin{table*}
\centering
\caption[]{Journal of Observations}
\begin{tabular} {l c c c c c c}
\hline
 Date & Spectral range &       Disp.          & FWHM & Spatial res.            & PA   & Exposure Time \\
         &     (\AA)          & (\AA\,px$^{-1}$) &  (\AA)   & (\arcsec\,px$^{-1}$) &  ($ ^{o} $) & (sec)   \\
\hline
04-02-2000 & 4779-5199  &       0.21       &  0.4 &   0.38    &    355    &  4\,$\times$\,1200 \\
04-02-2000 & 8363-8763  &       0.39       &  0.7 &   0.36   &    355    &  4\,$\times$\,1200  \\
05-02-2000 & 4779-5199  &       0.21       &  0.4 &   0.38    &    45    &   3\,$\times$\,1200 \\
05-02-2000 & 8363-8763  &       0.39       &  0.7 &   0.36   &    45    &   3\,$\times$\,1200 \\
05-02-2000 & 4779-5199  &       0.21       &  0.4 &   0.38    &    310   &   3\,$\times$\,1200 \\
05-02-2000 & 8363-8763  &       0.39       &  0.7 &   0.36   &    310    &  3\,$\times$\,1200  \\
\hline
\end{tabular}
\label{journal}
\end{table*}


High resolution blue and far-red spectra were acquired as part of an observing
run in 2000. They were obtained simultaneously using the blue and red arms of
the Intermediate dispersion Spectrograph and Imaging System (ISIS), on the
4.2\,m William Herschel Telescope (WHT) of the Isaac Newton Group (ING) at the
Roque de los Muchachos Observatory, on the Spanish island of La Palma. We used
the EEV12 detector in the blue arm with a CCD binning factor of 2 both
in ``x'' and
``y''.   The H2400B grating was used to cover the wavelength range
4779-5199\,\AA\ ($\lambda_c$\,=\,4989\,\AA). The TEK4 CCD was attached to
the red arm.  In this case, the R1200R grating was used, covering a spectral
range from 8363 to 8763\,\AA\ ($\lambda_c$\,=\,8563\,\AA). The spatial
resolution of the observations was 0.2 and 0.36\arcsec\,px$^{-1}$ for the
blue and red configurations respectively. A slit width of 1 arcsec was used.
The spectral ranges and grating resolutions (spectral dispersion in
\AA\,px$^{-1}$ and FWHM in \AA\ measured on the sky lines) attained are given
in table \ref{journal} containing the journal of observations.

\begin{figure*}
    \centering
    \includegraphics[width=0.48\textwidth]{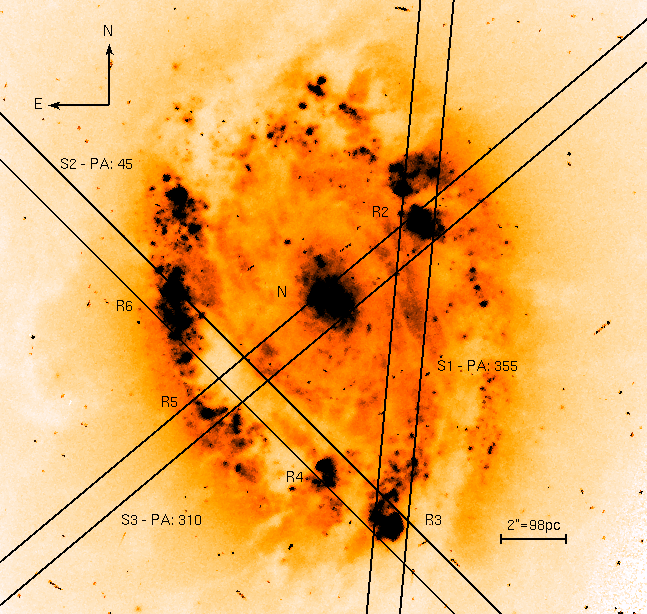}
\hspace{0.2cm}
\includegraphics[width=0.48\textwidth]{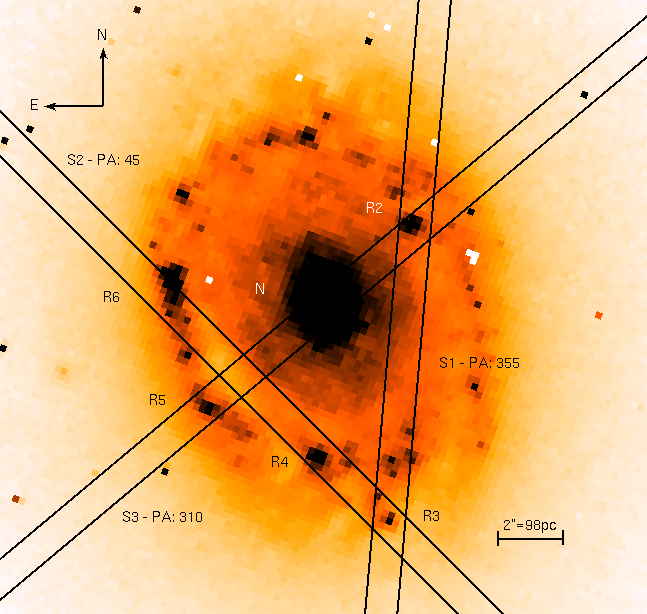}
\caption[]{Left: F606W (wide V) image centred on NGC\,3351 obtained with the WFPC2
  camera (PC1) of the Hubble Space Telescope. Right: HST-NICMOS image obtained through the F160W filter. For both images the orientation is north up, east to the left. The location and P.A. of the WHT-ISIS slit positions, together with
 identifications of the CNSFRs extracted, are marked.}
\label{hst-slits}
\end{figure*}

Three different slit positions were chosen to observe 5 CNSFRs and the nucleus
of NGC\,3351. In Fig.\ \ref{hst-slits} we show the position of these slits at
angles of 355$^{\circ}$, 45$^{\circ}$ and 310$^{\circ}$ (which we have 
labelled S1, S2 and S3
respectively) overlapped on an image of the circumnuclear region of NGC\,3351
acquired by the HST (see description below). The regions observed are
identified on the H$\alpha$ map by Planesas, Colina \& P\'erez-Olea (1997) using their same
nomenclature. The slit was positioned on the regions using the offsets from the nucleus, clearly visible during the data acquisition, given by  Planesas et al. (1997).

Several bias and sky flat field frames were taken at the beginning and the end
of each night in both arms. In addition, two lamp flat field and one 
calibration
lamp exposure per each telescope position were performed. The calibration lamp
used was CuNe+CuAr.

We have also downloaded two astrometrically and photometrically calibrated
broad-band images of the central part of NGC\,3351 from the Multimission Archive at Space 
Telescope\footnote{http://archive.stsci.edu/hst/wfpc2}. The images were taken
through the F606W (wide V)  and the F160W (H) filters with the
Wide Field and Planetary Camera 2 (WFPC2; PC1) and the NICMOS Camera 3 both on-board the HST. They are  
displayed in Figure 1,  with the slit positions overlaid and the CNSFRs and the nucleus labelled.

\subsection{Data reduction}

The data was processed and analysed using IRAF\footnote{IRAF: the Image
Reduction and Analysis Facility is distributed by the National Optical
Astronomy Observatories, which is operated by the Association of Universities
for Research in Astronomy, Inc. (AURA) under cooperative agreement with the
National Science Foundation (NSF).} routines in the usual manner.
The procedure includes the removal of cosmic
rays, bias substraction, division by a normalised flat field and wavelength
calibration. Wavelength fits were performed using 20-25 arc lines in
the blue and 10-15 lines in the far red by a polynomial of second- to
third-order. These fits have been done at 50 and 60 locations along the slit
in the blue and far red, respectively, and they have yielded rms residuals
between $\sim$0.1 and $\sim$0.2\,px. 

In the red, we also performed the wavelength calibration 
using sky lines following the work by \cite{1996PASP..108..277O}. However,
although  more lines (20-25) were available for the calibration, the fits gave
higher rms residuals, between $\sim$0.3 and $\sim$0.45\,px since the low
intensities of some of the lines did not allow a good Gaussian fit. We have
therefore adopted the calibration made using only the arc lines. 

Background subtraction was performed using the spectra at both ends of the
slit. This background includes  light from the disk and bulge of the galaxy. 
It was almost impossible to neatly substract the background bright emission
lines in the far red spectra due to the nonuniform and extended nebular
emission surrounding each cluster and the variation over time 
of the sky emission lines. It is worth noting that these spurious features do
not affect the CaT absorption lines.

We have not corrected the spectra for atmospheric extinction or
performed any flux calibration, since our purpose was to measure radial
velocities and velocity dispersions. 

In addition to the galaxy frames, observations of 11 template velocity stars
were made (4 during the first night and 7 during the second) to provide good
stellar reference frames in the same system as the galaxy spectra for the
kinematic analysis in the far red. They are late type giant and supergiant
stars which have strong CaT features \citep[see][]{1989MNRAS.239..325D}. In
table \ref{templates} we list the spectral types and luminosity classes of the
stars and the dates of observation.

\begin{table}
\centering
\caption[]{Stellar reference frames.}
\begin{tabular} {l l l}
\hline
Star      & ST\,-\,LC & date   \\
\hline
HD71952  &   K0\,I     &    04-02-2000   \\
HD129972 &   G6\,I     &       \\
HD134047 &   G6\,III   &       \\
HD144063 &   G4\,III   &       \\
\hline
HD16400 & G5\,III     &    05-02-2000  \\
HD22007  & G5\,I       &   \\
HD22156  & G6\,III     & \\
HD92588  &  K1\,I       & \\
HD102165 & F7\,I       & \\
HD115004 &  G8\,III     & \\
HD116365 &  K3\,III     & \\
\hline
\end{tabular}
\label{templates}
\end{table}

\begin{figure*}
\includegraphics[width=.41\textwidth,angle=0]{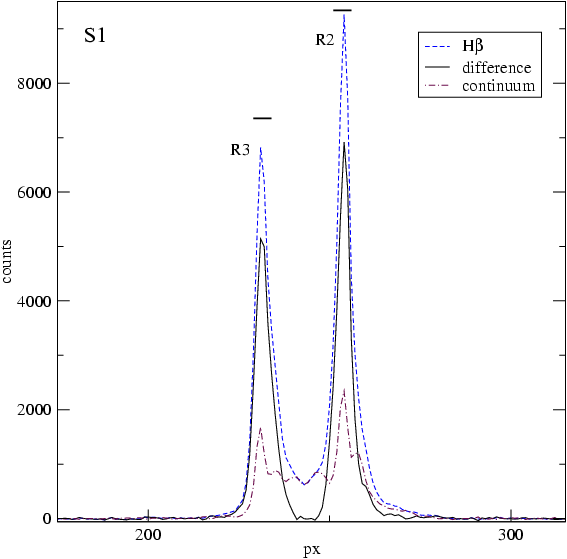}
\hspace{0.2cm}
\includegraphics[width=.41\textwidth,angle=0]{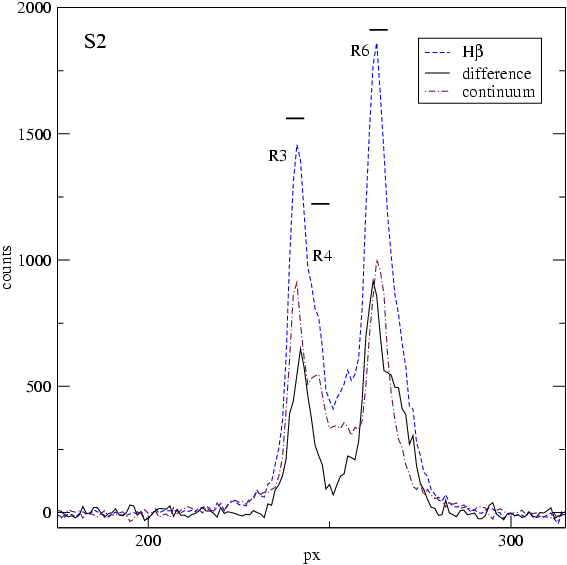}\\
\vspace{0.4cm}
\includegraphics[width=.41\textwidth,angle=0]{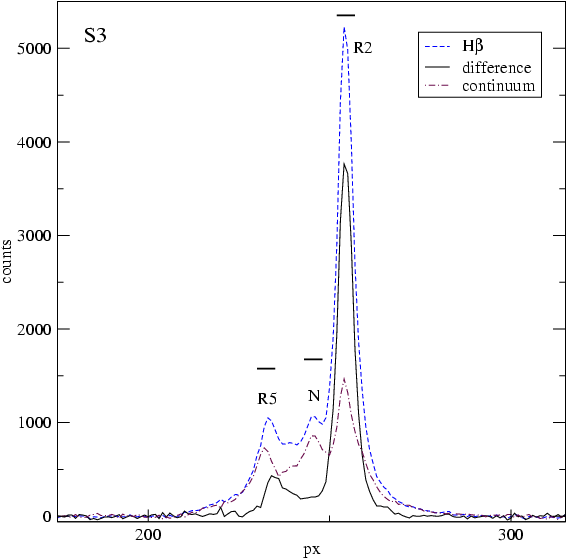}
\caption{Spatial profiles of H$\beta$ for each slit corresponding to
  line+continuum (dashed line), continuum (dashed-dotted line) and the
  difference between them (solid line), representing the pure emission from
  H$\beta$. Pixel number increases to the North. Horizontal small lines
  show the location of the CNSFRs and nuclear apertures.} 
\label{profiles}
\end{figure*}

Figure \ref{profiles} shows the spatial profiles of the H$\beta$ emission
along each slit position.  
These H$\beta$ profiles  have been generated by collapsing eleven pixels of
the spectra in the direction of the resolution between 4860 and 4862 \AA ,
rest frame,  $\lambda$ 4861 being the center of the H$\beta$ line, and are
plotted as dashed lines. Continuum profiles were generated by collapsing 11
resolution pixels centered on 4844 \AA\ and are plotted as dash-dotted
lines. The difference betwee the two is shown by solid lines and corresponds
to the pure H$\beta$ emission. 
On these profiles we have selected the regions of the frames to be extracted
in one dimensional spectra 
corresponding to each of the identified CNSFRs. Those regions are marked by
horizontal lines and labelled  
with their respective names in the figure. Spectra in slit positions 1 and
2 are placed along the circumnuclear ring and therefore any contribution from
the underlying galaxy bulge is difficult to assess but the spatial profiles
corresponding to slit position 3, that goes through the galactic nucleus, can
be used to make an estimate. For the blue spectra the light from the
underlying bulge is almost negligible amounting at most to 5\,\% at the
H$\beta$ 
line. For the red spectra, the contribution is more important. From Gaussian
fits 
to the $\lambda$\,8500\,\AA\ profile at slit position S3 we find that it
amounts to about 20\,\% for the 
lowest surface brightness region, R5. Similar values are found from inspection
of the HST-NICMOS image shown in Figure 1.

\begin{figure*}
\hspace{0.2cm}
\includegraphics[width=.63\textwidth,height=.31\textwidth,angle=0]{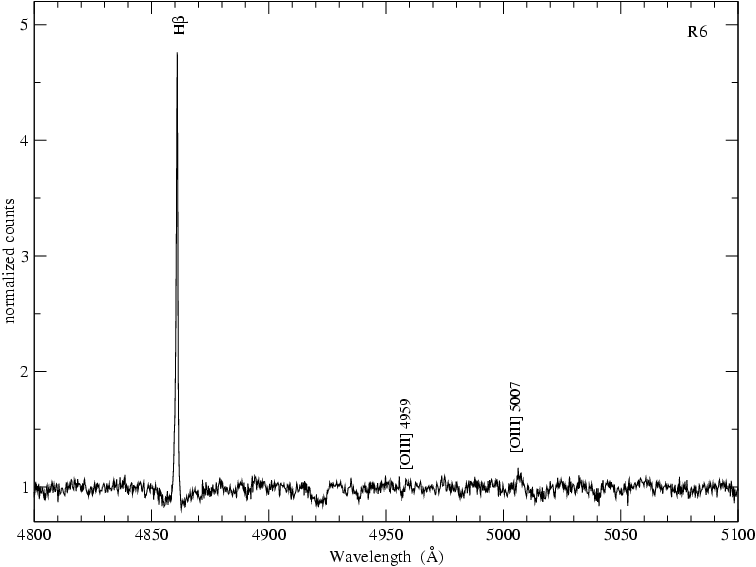}\\
\vspace{0.2cm}
\includegraphics[width=.63\textwidth,height=.31\textwidth,angle=0]{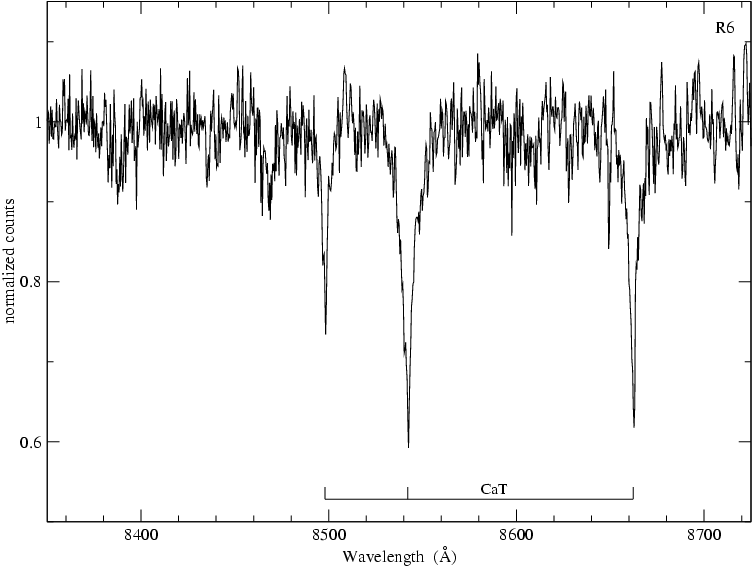}
\caption{Blue (upper panel) and red (lower panel) rest frame normalised
  spectra of R6. Notice the absence of conspicuous emission lines in the  
  red spectral range for this region.} 
\label{spectra-R6}
\end{figure*}

\begin{figure*}
\hspace{0.4cm}
\includegraphics[width=.64\textwidth,height=.31\textwidth,angle=0]{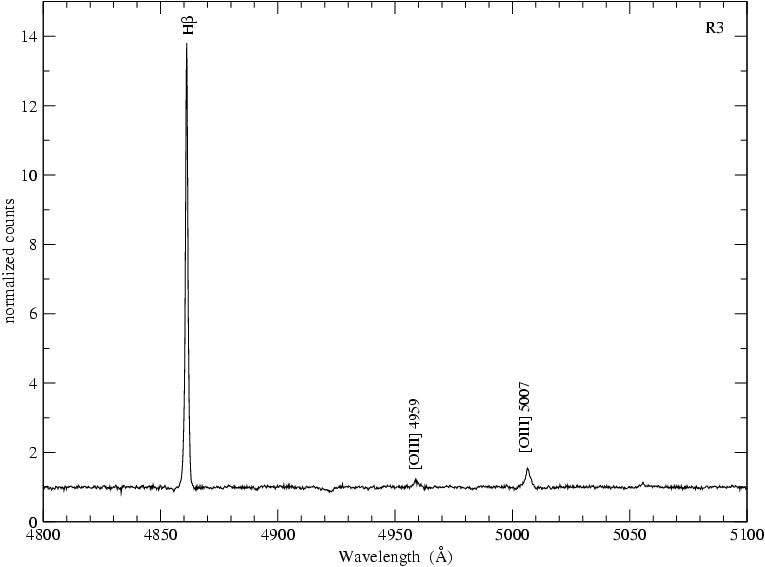}\\
\vspace{0.2cm}
\includegraphics[width=.63\textwidth,height=.31\textwidth,angle=0]{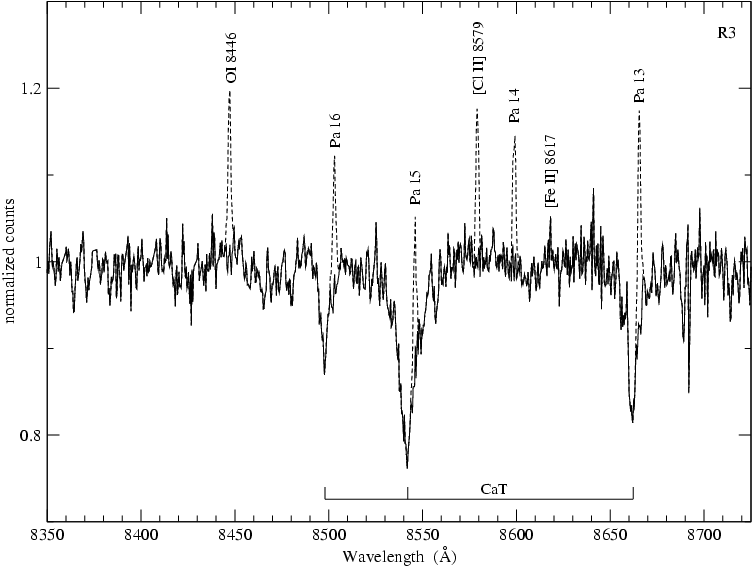}
\caption{Blue (upper panel) and red (lower panel) rest frame 
normalized spectra of R3. 
  The dashed line in the lower panel shows the obtained spectrum;
  the solid line represents the spectrum after subtracting the emission lines
  (see text).}
\label{spectra-R3}
\end{figure*}

Figs.\ \ref{spectra-R6} and \ref{spectra-R3} show representative spectra of
these regions (R6 and R3, respectively) in the blue and the red spectral range
split into two panels. The blue spectra 
show the Balmer H$\beta$ recombination line and the collisionally excited
[O{\sc iii}] lines at $\lambda\lambda$\,4959,5007\,\AA. Due to the high
metallicity of the CNSFRs  (D\'iaz et al.~2006), the lines of oxygen are seen
to be very weak (see 
Fig. \ref{enlarg}) and, in some cases, only the strongest $\lambda$ 5007 \AA\
is detected (left panel of this figure). The red spectra show the stellar
Ca{\sc ii} triplet (CaT) lines in absorption at $\lambda\lambda$ 8498, 8542,
8562\,\AA. In  some cases, these lines are contaminated by Paschen emission
(see for example the lower panel of Fig. \ref{spectra-R3}) which occur at
wavelengths very close to those of the CaT lines. In addition to these
emission lines, it is possible to observe other emission features,
such as O{\sc i}\,$\lambda$\,8446, [Cl{\sc ii}]\,$\lambda$\,8579, Pa\,14 and
[Fe{\sc ii}]\,$\lambda$\,8617. The observed red spectrum of R3 is plotted in
Fig.\ \ref{spectra-R3} with a dashed line. A single Gaussian fit to the
emission lines was performed and the lines were subsequently subtracted after
checking that the theoretically expected ratio between the Paschen lines was
satisfied. The solid line shows the subtracted spectrum.

\section{Results}
\label{Results}

\subsection{Kinematics of stars and ionised gas} 
\label{Method}

\subsubsection*{Stars}

Stellar radial velocities and velocity dispersions were obtained
from the absorption CaT lines using the cross-correlation technique  described
in detail by \cite{1979AJ.....84.1511T}. 
This method requires the comparison with a stellar template (which
can be synthetic or observed)  that represents the stellar
population that best reproduces the conspicuous feature used to perform these
measurements. An example of the red spectrum of a template star (HD116365) 
is shown in Fig. \ref{temspec} with the prominent features of CaT indicated.
The line-of-sight velocity dispersions is calculated from the width of
the primary peak of the cross-correlation function (CCF)
after deconvolution of the instrumental profile. A filtering of high
frequencies of the Fourier transform spectrum is usually included in this
procedure to avoid noise contamination and a low frequency filtering is
usually made to eliminate the residual continuum.

Minor changes/improvements with respect to the cross-correlation 
technique originally proposed by \cite{1979AJ.....84.1511T} were introduced as
described below. In order to apply this
method, we have used XCSAO, an external package of IRAF within the RVSAO,
which implements the cross-correlation method of Tonry and Davies and is a
direct descendant of the system built by them \citep{1998PASP..110..934K}.
We used late type giant and
supergiant stars that have strong CaT absorption lines (Fig.\ \ref{temspec})
as stellar velocity templates.
We normalized the stellar spectra dividing by a fitted continuum and convolved each stellar spectrum template with a set of Gaussian functions of different $\sigma$ simulating a wide range in velocity dispersions from 10 to 100 km/s with a bin size of 5 km/s. The obtained spectra are cross-correlated with the original template obtaining a relation between the width of the main peak of the cross-correlation and $\sigma$ of the input Gaussian. This relation constitutes a correction curve for each template that is applied as described below to obtain the stellar velocity dispersion for each CNSFR as described in \cite{1995ApJS...99...67N}. This procedure will allow us to correct for the known possible mismatches between template stars and the region composite spectrum. In Figure 7  we show an example of these correction curves, in particular for HD 144063, together with a linear fit to it.

\begin{figure*}
\includegraphics[width=.48\textwidth,height=.30\textwidth,angle=0]{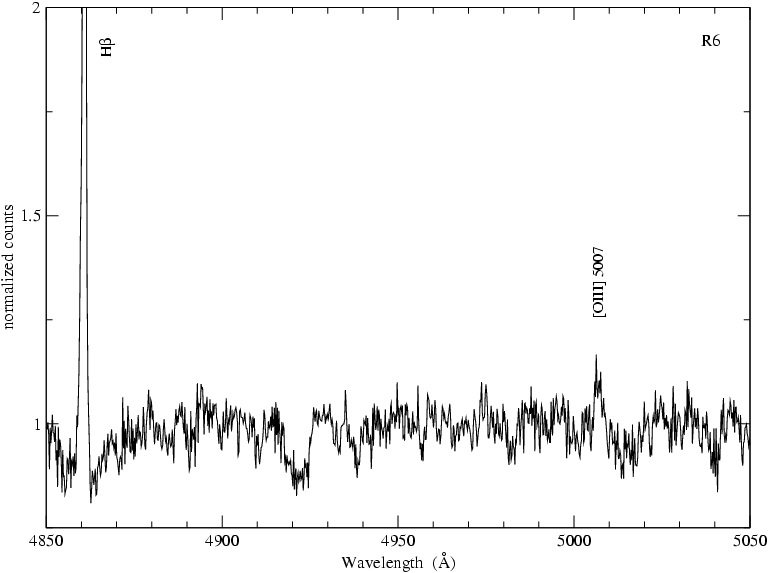}
\includegraphics[width=.48\textwidth,height=.30\textwidth,angle=0]{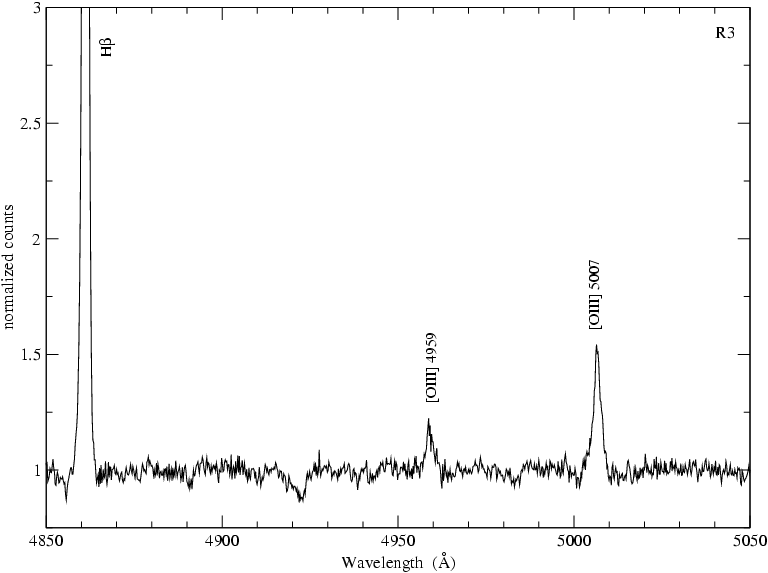}
\caption{Enlargement of the blue rest frame normalised spectra of R6 (left) and R3
  (right).}
\label{enlarg}
\end{figure*}

We determined the line of sight stellar velocity  along each
slit. Extractions were made  every two pixels for slit position S1. For slit
positions S2 and S3 they were made every three pixels, with one pixel overlap
between consecutive extractions. In this way 
the signal to noise ratio and the spatial
resolution were optimised. 

The stellar velocity dispersion was estimated at the
position of each CNSFR and the nucleus using apertures of 5 pixels in all
cases, which correspond to 1.0\arcsec\,$\times$\,1.8\arcsec .   A set of
eleven templates of different spectral types and 
luminosity classes were used following  \cite{1995ApJS...99...67N}
with the variation introduced by \cite{1997A&A...323..749P} of using the
individual stellar templates instead of an average. To measure the velocity
dispersion in a galactic spectrum we have convolved it with each stellar
template, correcting the width of the main peak of the CCF (see Fig.\
\ref{cross}) with the corresponding correction curve. Although the linear fits
to the curves are very good approximations (see Fig.\ \ref{correction}), we
used a linear interpolation between the two nearest values to estimate
the corrected width. The $\sigma$ of the stars ($\sigma_{\ast}$) is the
average of the $\sigma$ values 
found for each stellar template and its error is given by the dispersion of
the individual values of $\sigma$ and the rms of the residuals of the
wavelength fit. This procedure allows a more accurate
estimate of $\sigma_{\ast}$ \citep{1997A&A...323..749P}.
The radial velocities are the average of the radial velocities determined
directly from the 
position of the main peak of the cross-correlation of each galaxy spectrum
with each template in the rest frame.

\begin{figure*}
\includegraphics[width=.63\textwidth,height=.31\textwidth,angle=0]{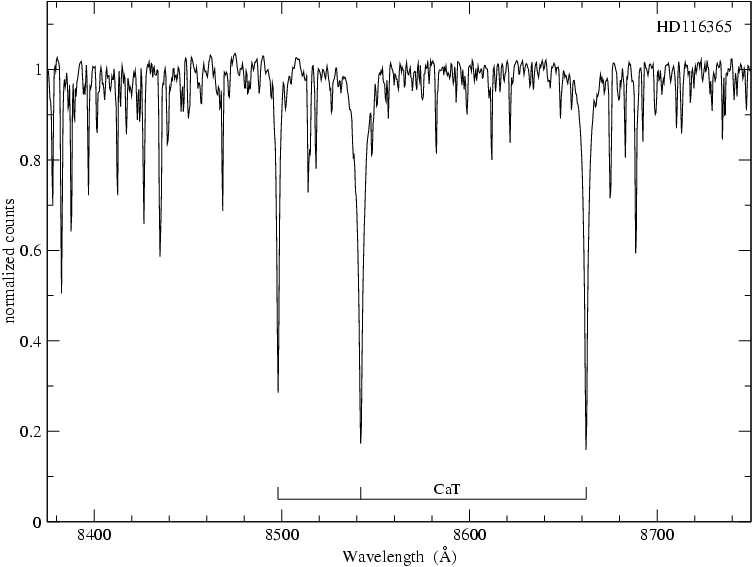}
\caption{Red rest frame normalised spectrum of HD116365.}
\label{temspec}
\end{figure*}

\begin{figure}
\centering
\includegraphics[width=.35\textwidth]{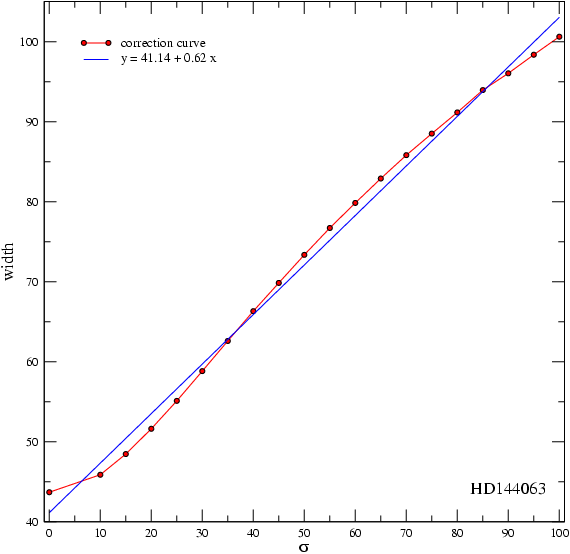}
\caption{Velocity dispersions correction curve for HD144063 (circles). The
  solid line is a linear fit to the curve.}
\label{correction}
\end{figure}

The stellar velocity dispersions are listed in column 3 of table
\ref{disp} along with their corresponding errors.

\subsubsection*{Ionised gas}

The wavelength and the width of the H$\beta$ and [O{\sc
iii}]\,$\lambda$\,5007\,\AA\ emission lines were measured to determine both
the radial velocities and the velocity dispersions of the ionised gas. As in
the case of the stars, we determined the radial velocity of the gas in the
line of 
sight along each slit, every two pixels for S1 and every three pixels,
superposing one pixel for consecutive extractions, for S2 and S3. 

The velocity dispersion of the gas was estimated at the position of each CNSFR
and the nucleus using 
5 pixel apertures, corresponding to 1.0\arcsec\,$\times$\,1.9\arcsec.
Following \cite{2000MNRAS.317..907J} we adjusted three
different suitable continua chosen by visual inspection and fitted a
Gaussian to the whole line. Positions and widths of the emission lines are the
average of the corresponding measurements and their errors are
calculated as the dispersion of these measurements taking into account the rms
of the residuals 
of the wavelength calibration. Thus, the error is associated with the
continuum placement. 

\begin{figure}
\includegraphics[width=.45\textwidth,height=.30\textwidth]{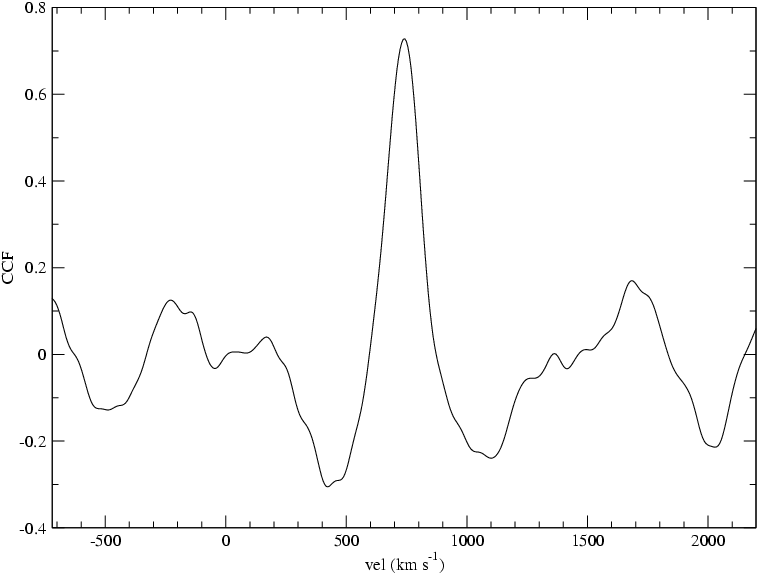}
\caption{Example of the cross correlation function.}
\label{cross}
\end{figure}

The velocity dispersions of the gas are calculated as
\[
\sigma_{gas}\,=\,\sqrt{\sigma_m^2\,-\,\sigma_i^2}
\]
\noindent where $\sigma_m$ and $\sigma_i$ are the measured and 
instrumental dispersions respectively. $\sigma_i$ was  measured directly from
the sky emission lines and is about 10.5\,km\,s$^{-1}$ at $\lambda$\,4861\,\AA.

\begin{figure*}
\includegraphics[width=.48\textwidth,height=.30\textwidth,angle=0]{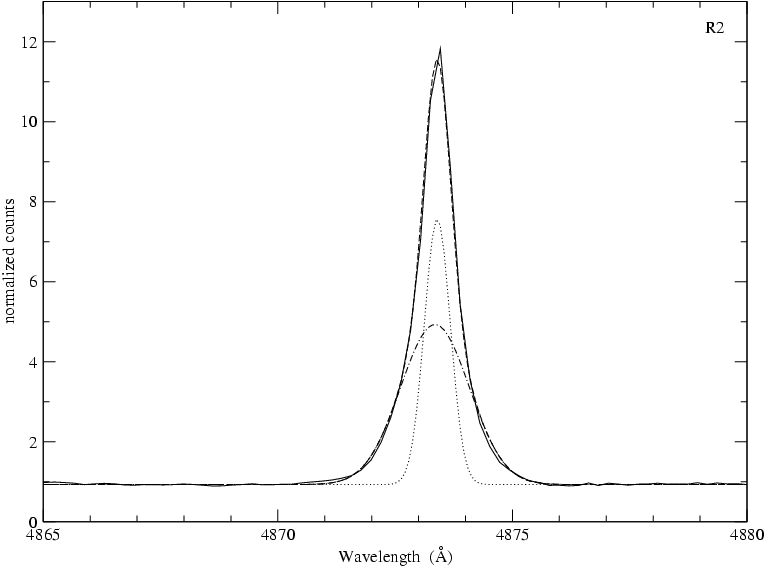}
\includegraphics[width=.48\textwidth,height=.30\textwidth,angle=0]{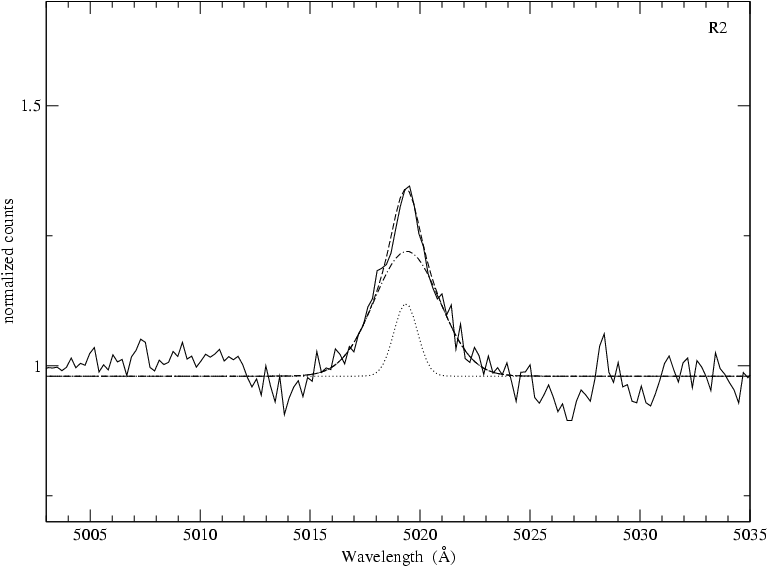}
\caption{Sections of the normalised spectrum of R2. The left panel shows
 from 4865 to 4880\,\AA, containing  H$\beta$ and the right panel,
  from 5003 to 5035\,\AA, containing the [O{\sc
  iii}]\,$\lambda$\,5007\,\AA\ emission line. For both  we have
  superposed the fits from the ngaussfit task in IRAF;
  the dashed-dotted line is the broad component, the dotted line is the
  narrow component and the dashed line is the sum of both.} 
\label{ngauss}
\end{figure*}

Unexpectedly, the Gaussian fit just described revealed the presence of more
than one component in the H$\beta$ lines. The optimal fit was found for two
different components for all the regions. 
We then used the widths of those components to fit the [OIII] lines finding
also optimal fits for regions R2 and R3 in the four analysed spectra
corresponding to two different slit positions.  For the rest of the regions
the two component fit did not show any improvemt over the single component
one, probably due to the low signal-to-noise ratio. An example of the fit for
H$\beta$ and [O{\sc iii}]\,$\lambda$\,5007\,\AA\ can be seen in 
Fig. \ref{ngauss}. The radial velocities found using this method are the same,
within 
the errors, as those found by fitting a single component. 

Gas velocity dispersions for H$\beta$ and [O{\sc
    iii}]\,$\lambda$\,5007\,\AA\ lines are given in table \ref{disp}. We have
listed the velocity dispersions derived using the two different methods, the
single line and the double line Gaussian fit . For the first method we have
labelled the results as {\it 1 component} (columns 4 and 5). For the second
one, labels {\it 2 components - narrow} (columns 6 and 7) and {\it 2
components - broad} (columns 8 and 9) for the narrow and the  broad component
respectively are used.  



\begin{table*}
\centering
\caption[]{Velocity dispersions.}
\begin{tabular} {l c c c c c c c c}
\hline
        &      &                  &  \multicolumn{2}{c}{{\it 1 component}} &
        \multicolumn{4}{c}{{\it 2 components}}    \\
        &      &           &       &      &  \multicolumn{2}{c}{{\it narrow}}
        & \multicolumn{2}{c}{{\it broad}}  \\
 Region & Slit & $\sigma_{\ast}$  & $\sigma_{gas}$(H$\beta$) &
 $\sigma_{gas}$([O{\sc iii}]) &  $\sigma_{gas}$(H$\beta$)   &
        $\sigma_{gas}$([O{\sc iii}]) & $\sigma_{gas}$(H$\beta$)  &
        $\sigma_{gas}$([O{\sc iii}]) \\
\hline

R2  &  S1   &  50$\pm$1   & 26$\pm$1  & 72$\pm$7 & 17$\pm$3  & 21$\pm$4  &  45$\pm$3 & 74$\pm$5  \\
R2  &  S3   &  51$\pm$6   & 29$\pm$3  & 69$\pm$9 & 16$\pm$2  & 23$\pm$5  &  43$\pm$2 & 76$\pm$8  \\
R3  &  S1   &  55$\pm$5   & 35$\pm$1  & 67$\pm$7 & 25$\pm$3  &  28$\pm$4 &  59$\pm$4 & 71$\pm$4  \\
R3  &  S2   &  59$\pm$7   & 39$\pm$5  & 70$\pm$7 & 24$\pm$3  &  24$\pm$6 &  59$\pm$3 & 74$\pm$9  \\
R4  &  S2   &  66$\pm$4   & 37$\pm$4  & 76$\pm$8 & 29$\pm$3  &  ---      &  65$\pm$4 &   ---     \\
R5  &  S3   &  47$\pm$4   & 34$\pm$2  & 56$\pm$7 & 30        &  ---      &  76       & ---       \\
R6  &  S2   &  39$\pm$6   & 29$\pm$6  & 46$\pm$7 & 16$\pm$3  &  ---      &  46$\pm$4 & ---       \\[2pt]
N   &  S3   &  67$\pm$1   & 53$\pm$3  & 73$\pm$6 & 41$\pm$5  &  ---      &  67$\pm$7 & ---       \\

\hline
\multicolumn{9}{l}{velocity dispersions in km\,s$^{-1}$} \\

\end{tabular}
\label{disp}
\end{table*}


\begin{figure*}
\includegraphics[width=.48\textwidth,angle=0]{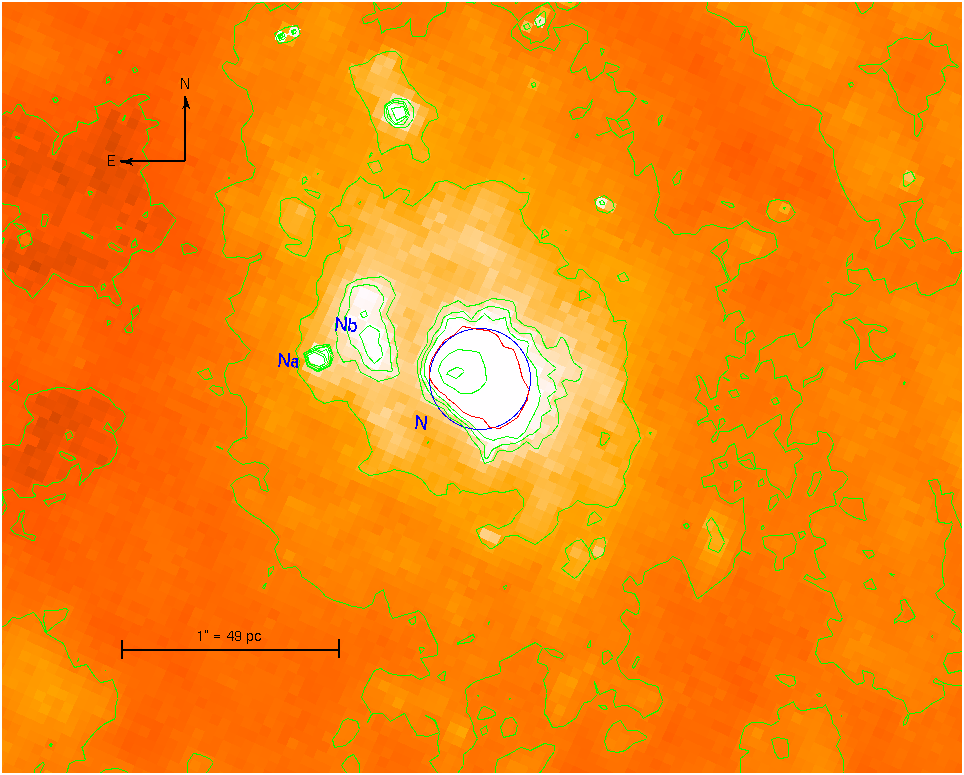}
\includegraphics[width=.48\textwidth,angle=0]{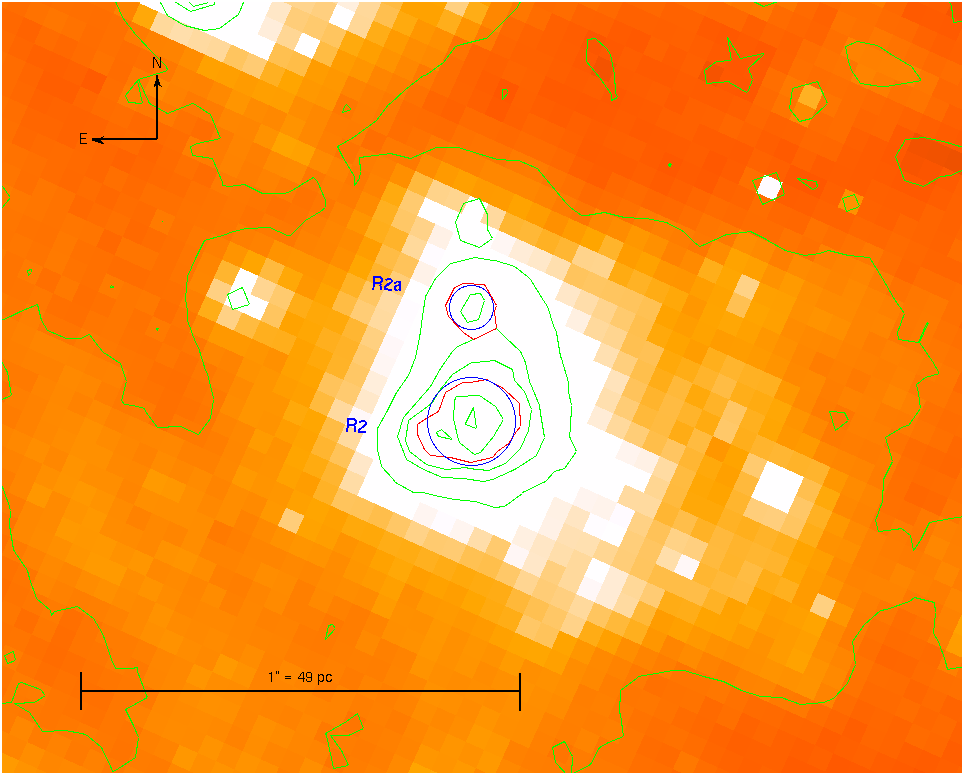}\\
\includegraphics[width=.48\textwidth,angle=0]{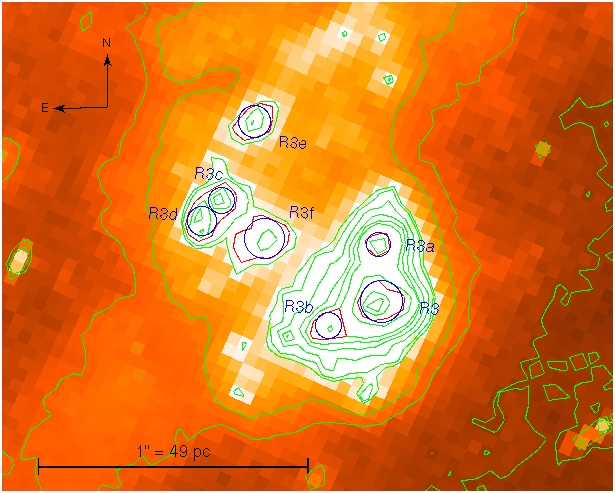}
\includegraphics[width=.48\textwidth,angle=0]{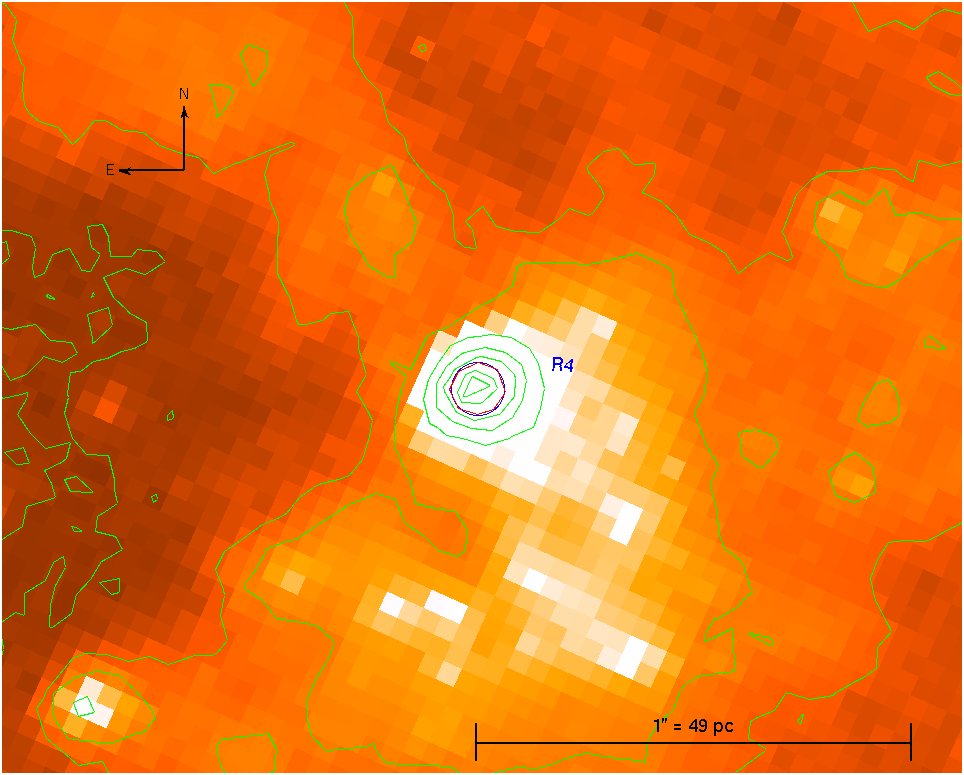}\\
\includegraphics[width=.48\textwidth,angle=0]{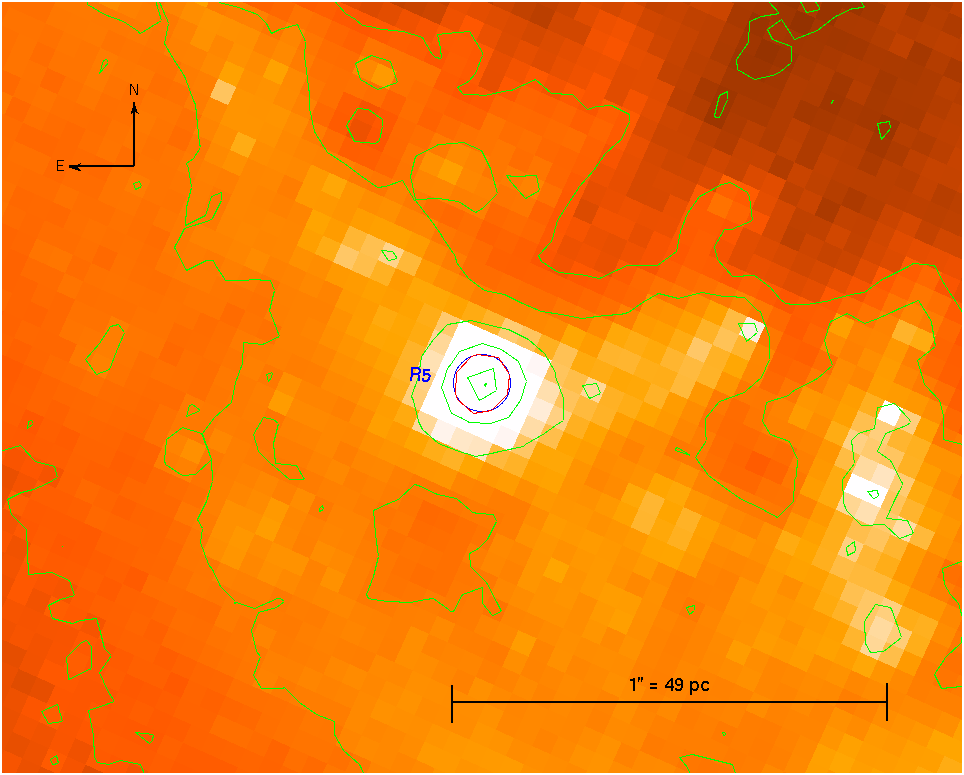}
\includegraphics[width=.48\textwidth,angle=0]{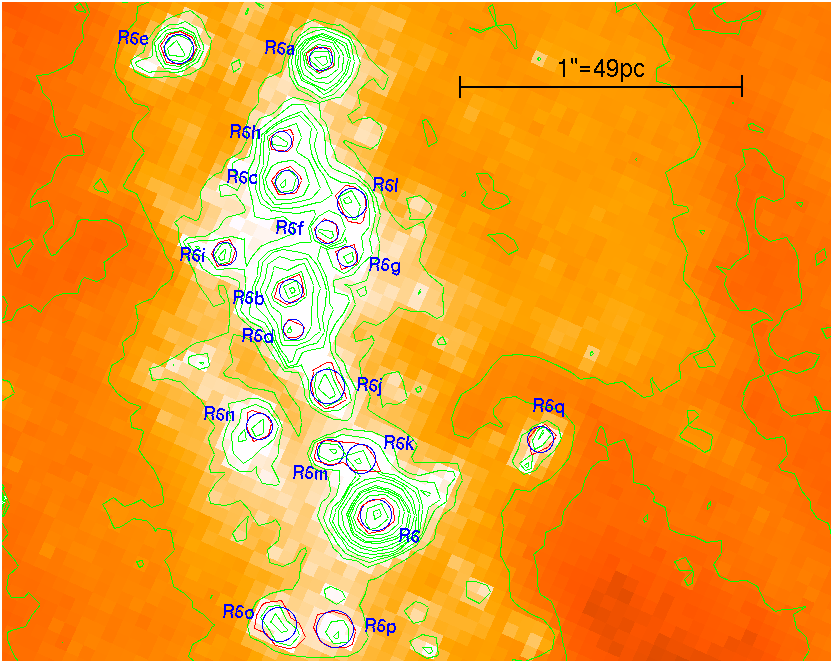}
\caption{Enlargements of the F606W image around the CNSFRs of our study with
  the contours overlapped. The circles correspond to the adopted radius
  for each region. [{\it See the electronic edition of the Journal for a  
  colour version of this figure where the adopted radii are in blue and
  the contours corresponding to the half light brightness are in red.}]} 
\label{sizes}
\end{figure*}

\subsection{Sizes}

Two parameters are needed in order to determine the mass of a virialized
stellar system, namely its velocity dispersion and its size.
Previously estimated radii (R) for the regions, such as those given
by Planesas et al.\ (1997)  from H$\alpha$ images or \cite{1997ApJ...484L..41C}
from ultraviolet STIS-HST images, were defined to
include the total integrated emission flux of the regions, and therefore they
are not appropriate to calculate their dynamical masses, as what is
needed for this is a measurement of the size of the mass distribution, i.e. the
star-cluster size. This is traditionally done measuring the effective radius
using images obtained in bands where the light is dominated by the stellar
contribution and where the contamination by gaseous emission is either small or
can be estimated and corrected.

Following  \cite{1996ApJ...466L..83H} we have defined the
radius (R) of a given structure as in \cite{1995AJ....110.2665M} assuming that
they have an intrinsically circularly symmetric Gaussian profile. As we can
see in Fig.\ \ref{sizes} this is a rather good 
approximation. We have determined the full width at half maximum, $W_{50}$,
by superposing the contours in the F606W WFPC2-HST image and determining the
half 
light strength between the peak of the intensity for each region and the
galaxy background (red contours in the electronic version of the paper). Then,
for this model, we take R as the half light radius, 
R\,=\,0.5$\times$W$_{50}$. While \cite{1995AJ....110.2665M} measured the sizes
from an H$\alpha$ image, we only have the F606W (wide V; containing H$\alpha$)
image in this spectral range and with the appropriate spatial resolution.

We find, as expected, that several of these regions are composed of
more than one knot. Only R4 and R5 seem to have a unique principal knot, at
least at the PC1 spatial resolution. In the most extreme case, R6, we find at
least 18 knots with a detection level of 10\,$\sigma$ above background
level. All these knots are within the radius of the regions defined by
Planesas et al.\ (1997).  We have to remark that our search for knots
has not been exhaustive since that is not the aim of this work. The radii of
the single knots vary between 1.7 and 4.9\,pc.

Position, radius and its error, and the peak intensity in counts measured from
the WFPC2 image are listed in table \ref{knots} for each knot identified.



\begin{table}
\centering
\caption[]{Positions, radii and peak intensities.}
\begin{tabular} {l c c c c }
\hline Region &  \multicolumn{2}{c}{position} &  R & I \\
  &  $\alpha_{J2000.0}$ & $\delta_{J2000.0}$ & (pc) & (counts)  \\
\hline

R2  & 10$^{\romn h}$43$^{\romn m}$57\fs12  &  +11$^\circ$42\arcmin08\farcs27  &
 4.9$\pm$0.2  & 2531  \\
R2a & 10$^{\romn h}$43$^{\romn m}$57\fs12  &  +11$^\circ$42\arcmin08\farcs53  &
 2.5$\pm$0.2  &  956  \\[2pt] 

R3  & 10$^{\romn h}$43$^{\romn m}$57\fs16  &  +11$^\circ$41\arcmin59\farcs02  &
 3.9$\pm$0.2  & 2598  \\
R3a & 10$^{\romn h}$43$^{\romn m}$57\fs16  &  +11$^\circ$41\arcmin59\farcs24  &
 2.2$\pm$0.1  & 1852  \\
R3b & 10$^{\romn h}$43$^{\romn m}$57\fs18  &  +11$^\circ$41\arcmin58\farcs93  &
 2.0$\pm$0.3  & 1032  \\
R3c & 10$^{\romn h}$43$^{\romn m}$57\fs21  &  +11$^\circ$41\arcmin59\farcs41  &
 2.5$\pm$0.3  &  546  \\
R3d & 10$^{\romn h}$43$^{\romn m}$57\fs21  &  +11$^\circ$41\arcmin59\farcs33  &
 2.7$\pm$0.3  &  489  \\
R3e & 10$^{\romn h}$43$^{\romn m}$57\fs20  &  +11$^\circ$41\arcmin59\farcs71  &
 2.9$\pm$0.2  &  383  \\
R3f & 10$^{\romn h}$43$^{\romn m}$57\fs20  &  +11$^\circ$41\arcmin59\farcs26  &
 3.7$\pm$0.4  &  330  \\[2pt]

R4  & 10$^{\romn h}$43$^{\romn m}$57\fs32  &  +11$^\circ$42\arcmin00\farcs92  &
 2.9$\pm$0.1  & 2581  \\[2pt]

R5  & 10$^{\romn h}$43$^{\romn m}$57\fs55  &  +11$^\circ$42\arcmin02\farcs50  &
 3.2$\pm$0.1  &  973  \\[2pt]

R6  & 10$^{\romn h}$43$^{\romn m}$57\fs60  &  +11$^\circ$42\arcmin05\farcs21  &
 2.7$\pm$0.1  & 2399  \\
R6a & 10$^{\romn h}$43$^{\romn m}$57\fs62  &  +11$^\circ$42\arcmin06\farcs81  &
 2.0$\pm$0.1  &  794  \\
R6b & 10$^{\romn h}$43$^{\romn m}$57\fs62  &  +11$^\circ$42\arcmin06\farcs00  &
 2.0$\pm$0.2  &  714  \\
R6c & 10$^{\romn h}$43$^{\romn m}$57\fs63  &  +11$^\circ$42\arcmin06\farcs38  &
 2.0$\pm$0.1  &  551  \\
R6d & 10$^{\romn h}$43$^{\romn m}$57\fs62  &  +11$^\circ$42\arcmin05\farcs86  &
 1.7$\pm$0.1  &  404  \\
R6e & 10$^{\romn h}$43$^{\romn m}$57\fs65  &  +11$^\circ$42\arcmin06\farcs85  &
 2.4$\pm$0.2  &  376  \\
R6f & 10$^{\romn h}$43$^{\romn m}$57\fs62  &  +11$^\circ$42\arcmin06\farcs21  &
 2.0$\pm$0.1  &  340  \\
R6g & 10$^{\romn h}$43$^{\romn m}$57\fs61  &  +11$^\circ$42\arcmin06\farcs12  &
 1.7$\pm$0.1  &  295  \\
R6h & 10$^{\romn h}$43$^{\romn m}$57\fs63  &  +11$^\circ$42\arcmin06\farcs52  &
 1.7$\pm$0.1  &  290  \\
R6i & 10$^{\romn h}$43$^{\romn m}$57\fs64  &  +11$^\circ$42\arcmin06\farcs13  &
 2.0$\pm$0.1  &  285  \\
R6j & 10$^{\romn h}$43$^{\romn m}$57\fs62  &  +11$^\circ$42\arcmin05\farcs66  &
 2.9$\pm$0.3  &  282  \\
R6k & 10$^{\romn h}$43$^{\romn m}$57\fs61  &  +11$^\circ$42\arcmin05\farcs41  &
 2.4$\pm$0.2  &  275  \\
R6l & 10$^{\romn h}$43$^{\romn m}$57\fs61  &  +11$^\circ$42\arcmin06\farcs31  &
 2.4$\pm$0.3  &  272  \\
R6m & 10$^{\romn h}$43$^{\romn m}$57\fs61  &  +11$^\circ$42\arcmin05\farcs43  &
 2.2$\pm$0.2  &  270  \\
R6n & 10$^{\romn h}$43$^{\romn m}$57\fs63  &  +11$^\circ$42\arcmin05\farcs52  &
 2.2$\pm$0.3  &  251  \\
R6o & 10$^{\romn h}$43$^{\romn m}$57\fs63  &  +11$^\circ$42\arcmin04\farcs83  &
 2.9$\pm$0.3  &  212  \\
R6p & 10$^{\romn h}$43$^{\romn m}$57\fs61  &  +11$^\circ$42\arcmin04\farcs81  &
 3.1$\pm$0.3  &  212  \\
R6q & 10$^{\romn h}$43$^{\romn m}$57\fs56  &  +11$^\circ$42\arcmin05\farcs48  &
 2.2$\pm$0.3  &  211  \\[2pt]

N   & 10$^{\romn h}$43$^{\romn m}$57\fs29  &  +11$^\circ$42\arcmin05\farcs81  &
 11.3$\pm$0.1  & 513  \\[2pt]

\hline

\end{tabular}
\label{knots}
\end{table}



\begin{table}
\centering
\caption[]{Dynamical masses.}
\begin{tabular} {l c c c }
\hline
 Region & Slit & M$_{\ast}$ & error(\%)  \\ 
  
\hline

R2  &  S1   &  85$\pm$5   &   \\
R2a &  S1   &  44$\pm$4   &  \\
R2sum & S1  & 129$\pm$6   & 3.4  \\[2pt]
R2  &  S3   &  87$\pm$21  &   \\ 
R2a &  S3   &  44$\pm$11  &  \\ 
R2sum & S3  & 131$\pm$23  & 17.8 \\[2pt] 

R2 (adopted) & & 129$\pm$6 & 3.4 \\[4pt] 

R3  &  S1   &  82$\pm$14  &   \\
R3a &  S1   &  46$\pm$8   &    \\
R3b &  S1   &  42$\pm$9   &    \\
R3c &  S1   &  52$\pm$11  &    \\
R3d &  S1   &  57$\pm$11  &   \\
R3e &  S1   &  61$\pm$11  &   \\
R3f &  S1   &  78$\pm$15  &   \\
R3sum & S1  & 417$\pm$31  & 7.3  \\[2pt]
R3  &  S2   &  94$\pm$23  &   \\ 
R3a &  S2   &  53$\pm$13  &  \\ 
R3b &  S2   &  48$\pm$14  &   \\ 
R3c &  S2   &  60$\pm$16  &   \\ 
R3d &  S2   &  65$\pm$17  &   \\ 
R3e &  S2   &  70$\pm$17  &   \\ 
R3f &  S2   &  89$\pm$5   &  \\ 
R3sum & S2  & 477$\pm$47  & 9.9  \\[2pt] 

R3 (adopted) & & 447$\pm$56 & 12.6 \\[4pt] 

R4  &  S2   &  87$\pm$12  & 13.9 \\[4pt] 

R5  &  S3   &  49$\pm$8   & 16.4  \\[4pt] 

R6  &  S2   &  29$\pm$9   &  \\
R6a &  S2   &  21$\pm$8   &   \\
R6b &  S2   &  21$\pm$8   &  \\
R6c &  S2   &  21$\pm$8   &  \\
R6d &  S2   &  18$\pm$7   &   \\
R6e &  S2   &  26$\pm$10  &   \\
R6f &  S2   &  21$\pm$8   &  \\
R6g &  S2   &  18$\pm$7   &  \\
R6h &  S2   &  18$\pm$7   &  \\
R6i &  S2   &  21$\pm$8   & \\
R6j &  S2   &  31$\pm$12  &  \\
R6k &  S2   &  26$\pm$10  &    \\
R6l &  S2   &  26$\pm$10  &  \\
R6m &  S2   &  24$\pm$9   &    \\
R6n &  S2   &  24$\pm$9   &   \\
R6o &  S2   &  31$\pm$12  &   \\
R6p &  S2   &  33$\pm$13  &   \\
R6q &  S2   &  24$\pm$9   &    \\
R6sum & S2  & 434$\pm$39  & 9.0  \\[4pt]

N   &  S3   & 350$\pm$11  & 3.1 \\[2pt]

\hline
\multicolumn{4}{l}{masses in 10$^5$ M$_\odot$} \\
\end{tabular}
\label{mass}
\end{table}


\subsection{Masses}

\label{masses}

\subsubsection*{Dynamical masses}

We have used the virial theorem to estimate upper limits to the dynamical
masses (M$_{\ast}$) inside the half light radius (R) for each observed knot in
the F606W WFPC2-HST image. In order to do this, we have 
assumed that the systems are spherically 
symmetric, gravitationally  bound and have isotropic velocity distribution
($\sigma^2$(total)\,=\,3 $\sigma_{\ast}^2$). Then, the dynamical mass is given
by M$_{\ast}$\,=\,3 $\sigma_{\ast}^2$ R/G
\citep{1996ApJ...466L..83H,1996ApJ...472..600H}.   

It must be noted that we have measurements for the
size of each knot (typically 5pc), but we do not have direct access
to the stellar velocity dispersion of each of the clusters, since
our spectroscopic measurements encompass a wider area
(1.0\arcsec\,$\times$\,1.8\arcsec\,$\sim$\,49\,$\times$\,88\,pc$^2$) that
includes the CNSFRs to which each group of knots belong. 
The use of these wider size scale velocity dispersion measurements to
estimate the mass of each knot, lead us to overestimate the mass of the
individual clusters, and hence of each CNSFR.

However, as can be seen in the HST-NICMOS image (right panel of Fig.
\ref{hst-slits}), the CNSFRs of NGC\,3351 are clearly visible in the IR and
dominate the light inside the apertures observed. All the regions analysed
show up very prominently in the near IR and therefore we can assume that the
light at the CaT wavelength region is dominated by the stars in the
clusters.  The IR CaII triplet is very strong, in fact the strongest stellar
feature, in very young clusters, i.e. older than 4\,Myr
\citep{1990MNRAS.242P..48T}. 
Besides, we detect a minimum in the velocity dispersion at the
position of the clusters, indicating that they are kinematically distinct. We
cannot be sure though that we are actually measuring their velocity
dispersion and thus prefer to say that our measurements of $\sigma_{\ast}$ and
hence dynamical masses constitute upper limits. Although we are well aware of
the difficulties, still we are confident that these upper limits are valid and
important for comparison with the gas kinematical measurements.

The estimated dynamical masses and their corresponding errors for each knot
are listed in table \ref{mass}. For the regions that have been observed in
more than one slit position, we list the derived values using the two
different stellar velocity dispersions. The dynamical masses in the lines
labelled  ``sum'' are the sum of the values derived for each knot of the
corresponding region, when there are more than one. The ``adopted''
dynamical mass for R2 is directly the sum of the dynamical masses of the two
knots derived using the velocity dispersion estimated from S1 since the mass
calculated from S3 is in agreement with the first one (see table
\ref{mass}) but it has a much greater error. This greater error is due to the
lower signal to noise ratio of the spectrum of R2 extracted from S3. 
In contrast, the derived dynamical masses for R3 are not in
complete agreement, although they coincide within the errors,
which are comparable. In this case we have taken  the
average  as the ``adopted'' value. The fractional error in the ``sum'' 
and the adopted values of the
dynamical masses of the CNSFRs are listed in column 4 of table
\ref{mass}.

\subsubsection*{Ionising star clusters}

We have derived the masses of the ionising star clusters (M$_{ion}$) from the
total number of ionising photons using solar metallicity single burst models
by \cite{1995A&AS..112...35G}, 
assuming a Salpeter IMF \citep{1955ApJ...121..161S} with lower and upper mass
limits of 0.8 and 120 
M$_\odot$ which provide the number of ionising photons per unit mass,
$(Q(H_0)/M_{ion})$. This number decreases with the age of the region. We have
used the following relation between $Q(H_0)/M_{ion}$ and the equivalent
width of H$\beta$, EW(H$\beta$), derived from the models, in order to take
into account the evolutionary state of the region
\citep{1998Ap&SS.263..143D}: 

\[
log\big(Q(H_0)/M_{ion}\big)\,=\,44.8\,+\,0.86\,log\big(EW(H\beta)\big)
\]

We have taken the values of EW(H$\beta$) from \cite{tesisdiego} 
(see table \ref{parameters}) which are not corrected by the contribution to
the continuum by the underlying stellar population. This correction would
increase the values of EW(H$\beta$) thus decreasing the calculated M$_{ion}$.
The total number of ionising photons has been derived from the H$\alpha$
luminosities \citep{1995ApJS...96....9L}:
\[
Q(H_0)\,=\,7.35\,\times\,10^{11}\,L(H\alpha)
\]

We have taken the total observed H$\alpha$ luminosities from Planesas et al.\
(1997) correcting them for the different assumed distance. These authors
estimated a diameter of 2.4\arcsec\ for each whole region and the nucleus,
except for R4 for which they used 2.2\arcsec. We  have corrected the H$\alpha$
luminosities for internal extinction using the colour excess [E(B-V)]
estimated by \cite{tesisdiego} and assuming  the  galactic extinction law of
\cite{1972ApJ...172..593M} with $R_v$\,=\,3.2.

Our derived values of Q(H$_0$) are lower limits since we have not taken
into account either the absorption of photons by dust or any photon escape
from the \HII\ regions.

The final expression for the derivation of M$_{ion}$ is: 

\[
M_{ion}\,=\,\frac{7.35\,\times\,10^{11}\,L(H\alpha)}{10^{44.8\,+\,0.86\,log\left[EW(H\beta)\right]}}
\]

\subsubsection*{Ionised gas}

The amount of ionised gas (M$_{{\rm HII}}$) associated to each  starforming
region complex has been obtained from our derived H$\alpha$ luminosities
using the relation given by \cite{1990ApJ...356..389M} for an electron
temperature of 10$^4$\,K 

\[
M_{{\rm HII}}\,=\,3.32\,\times\,10^{-33}\,L(H\alpha)\,N_e^{-1}
\]

\noindent where N$_e$ is the electron density. The electron density for each
region (obtained from  the [S{\sc ii}]\,$\lambda\lambda$\,6717\,/\,6731\,\AA\
line ratio) has been taken from D\'iaz et al.\ (2006) for the CNSFRs and
\cite{tesisdiego} for the nucleus (see table \ref{parameters}).

In table \ref{parameters} we have listed for each region and the nucleus, the
H$\alpha$ luminosities [L(H$\alpha$)] (corrected and uncorrected for
reddening), the colour excess, E(B-V), the logarithmic extinction at H$\alpha$,
C(H$\alpha$), the number of ionising photons, Q(H$_0$), the equivalent width
of H$\beta$, EW(H$\beta$), the electron density, N$_e$,  and the masses of the
ionising stellar cluster, M$_{ion}$, and of the ionised hydrogen, M$_{\rm
  HII}$.


\begin{table*}
\centering
\caption[]{Physical parameters.}
\begin{tabular} {l c c c c c c c c c c}
\hline
 Region & L$_{obs}$(H$\alpha$)$^{\dagger}$ & E(B-V)$^\ddagger$ &
 C(H$\alpha$) &L(H$\alpha$) & Q(H$_0$)  & 
 EW(H$\beta$)$^\ddagger$ & M$_{ion}$ & N$_e^\S$ & M$_{{\rm HII}}$ &
 M$_{ion}$/M$_{\ast}$(\%)
 \\ 
\hline

R2  & 19.3 & 0.17 &  0.02 & 20.5  & 15.0 &  9.5  &  7.2  & 440 & 0.15 &  5.6 \\
R3  & 25.0 & 0.46 &  0.65 & 111.0 & 81.5 &  16.5 &  24.2 & 430 & 0.86 &  5.4 \\
R4  & 12.7 & 0.27 &  0.24 & 22.0  & 16.2 &  13.0 &  5.9  & 310 & 0.24 &  6.8 \\
R5  & 5.9  & 0.25 &  0.20 & 9.2   & 6.8  &  5.1  &  5.5  & 360 & 0.09 &  11.3\\
R6  & 7.5  & 0.0  & -0.34 & 3.4   & 2.5  &  2.3  &  4.1  & 360 & 0.03 &  1.0 \\[2pt]
N   & 3.3  & 0.07 & -0.19 & 2.1   & 1.6  &  1.8  &  3.1  & 650$^\ddagger$ & 0.01
&  0.9 \\

\hline
\multicolumn{11}{l}{luminosities in 10$^{38}$\,erg\,s$^{-1}$, masses in 10$^5$
  M$_\odot$, ionising photons in 10$^{50}$\,ph\,s$^{-1}$ and densities in
cm$^{-3}$}\\
\multicolumn{11}{l}{$^\dagger$from Planesas et al.\ (1997) corrected for
 the different adopted distances} \\ 
\multicolumn{11}{l}{$^\ddagger$from P\'erez-Olea (1996)}\\
\multicolumn{11}{l}{$^{\S}$ from D\'iaz et al. (2006)}
\end{tabular}
\label{parameters}
\end{table*}

\subsection{Emission line ratios}

We have used two different ways to integrate the intensity of a given line:
(1) in the cases of a single Gaussian fit the  emission line intensities
were  measured using the SPLOT task of IRAF. The 
positions of the local continua are placed by eye. For the H$\beta$
emission lines a conspicuous underlying stellar population is easily
appreciable by the presence of absorption features that depress the lines
\cite[see discussion in][]{1988MNRAS.231...57D}. Examples of this effect can
be appreciated in Fig.\ \ref{enlarg}. We have then defined a pseudo-continuum
at the base  of the line to measure the line intensities and minimise
the errors introduced by the underlying population \cite[for details
  see][]{2006MNRAS.372..293H}. (2) in the cases of a fit by two
Gaussians the individual intensities  of the narrow and broad components
are estimated from the fitting parameters using the expression
I\,=\,1.0645\,A\,$\times$\,FWHM 
(=\,$\sqrt{2\pi}$\,A\,$\times$\,$\sigma$), where I is the Gaussian intensity,
A is the amplitude of the Gaussian and FWHM is the full width at half maximum
($\sigma$ is the dispersion of the Gaussian). A pseudo-continuum for 
the H$\beta$ emission line was also defined in these cases.

Following \cite{1994ApJ...437..239G}, \cite*{2002MNRAS.329..315C} and 
\cite{2003MNRAS.346..105P}, the statistical errors associated with the observed
emission fluxes have been calculated using the expression
$\sigma_{l}$\,=\,$\sigma_{c}$N$^{1/2}$[1 + EW/(N$\Delta$)]$^{1/2}$, where
$\sigma_{l}$ is  the error in the observed line flux, $\sigma_{c}$ represents
the standard 
deviation in a box near the measured emission line and stands for the error in
the continuum placement, N is the number of pixels used in the measurement of 
the line intensity, EW is the line equivalent width, and $\Delta$ is 
the wavelength
dispersion in angstroms per pixel. For the H$\beta$ emission 
line we have
doubled the derived error, $\sigma_{l}$, in order to take into account the
uncertainties introduced by the presence of the underlying stellar population 
\citep{2006MNRAS.372..293H}.

The logarithmic ratio between the emission line intensities of [O{\sc
    iii}]\,$\lambda$\,5007\,\AA\ and H$\beta$, and their corresponding errors,
are presented in table \ref{ratios}. We have also listed in this table the
logarithmic  ratio between the emission line fluxes of [N{\sc
    ii}]\,$\lambda$\,6584\,\AA\ and H$\alpha$ together with their
corresponding errors from \cite{2006astro.ph..0787D}.


\begin{table*}
\centering
\caption[]{Line ratios.}
\begin{tabular} {l c c c c c}
\hline
        &      &         {\it 1 component} &
        \multicolumn{2}{c}{{\it 2 components}}   & \\
        &      &    &  {\it narrow}  & {\it broad} &  \\
 Region & Slit & log([O{\sc iii}]5007/H$\beta$) & log([O{\sc
        iii}]5007/H$\beta$) & log([O{\sc iii}]5007/H$\beta$) & log([N{\sc
        ii}]6584/H$\beta$)$^{\dagger}$ \\
\hline

R2  &  S1   &  -1.07$\pm$0.06   & -1.66$\pm$0.08  & -0.93$\pm$0.12 & -0.43$\pm$0.01\\
R2  &  S3   &  -1.01$\pm$0.06   & -1.55$\pm$0.08  & -0.96$\pm$0.13 & \\
R3  &  S1   &  -1.10$\pm$0.06   & -1.57$\pm$0.07  & -0.93$\pm$0.10 & -0.42$\pm$0.01\\
R3  &  S2   &  -1.00$\pm$0.06   & -1.52$\pm$0.09  & -0.89$\pm$0.10 & \\
R4  &  S2   &  -1.03$\pm$0.07   &    ---          &        ---     & -0.49$\pm$0.01\\
R5  &  S3   &  -0.85$\pm$0.12   &    ---          &        ---     & -0.37$\pm$0.03\\
R6  &  S2   &  -1.09$\pm$0.11   &    ---          &        ---     & -0.52$\pm$0.02\\[2pt]
N   &  S3   &  -0.28$\pm$0.05   &    ---          &        ---     & \\

\hline

\multicolumn{6}{l}{$^{\dagger}$from D\'iaz et al.\ (2006)}

\end{tabular}
\label{ratios}
\end{table*}


\nocite{2003MNRAS.346.1055K}
\nocite{2001ApJ...556..121K}
\nocite{1997ApJS..112..315H}

\begin{figure}
\centering
\includegraphics[width=.44\textwidth,angle=0]{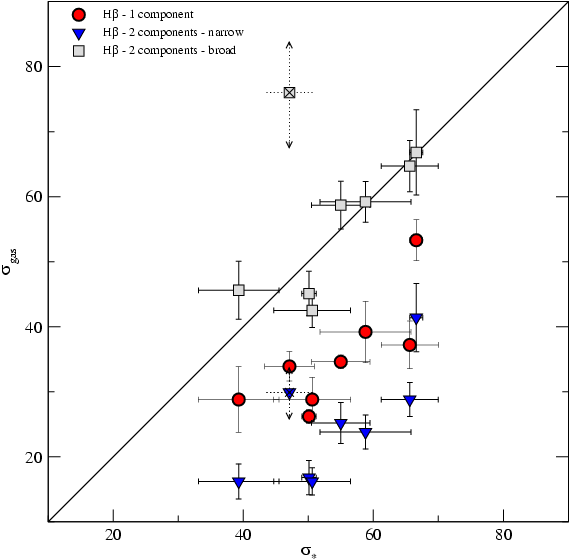}\vspace*{0.3cm}
\includegraphics[width=.44\textwidth,angle=0]{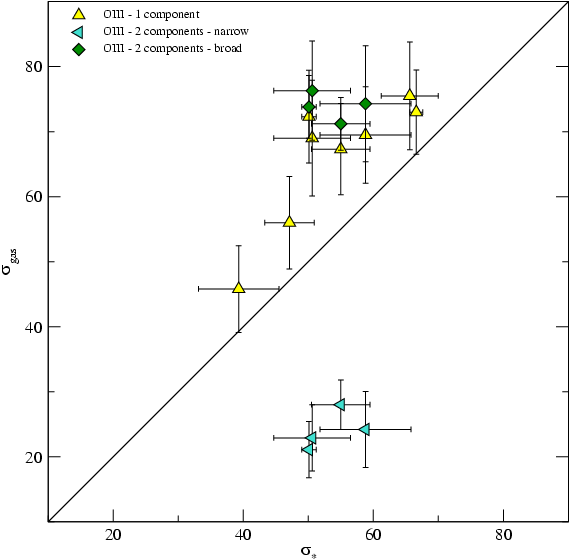}
\caption{Upper panel: relation between velocity dispersions of the gas (derived
  from H$\beta$) and stars (CaT) for the CNSFRs and the
  nucleus. Symbols are as follows: single Gaussian fit, circles;
  two Gaussian fit, broad component: squares; narrow component,
 triangles.
  Lower panel, as the upper panel for the [O{\sc
  iii}] line. Upward triangles  correspond to the estimates using a
  single Gaussian fit, diamonds represent the broad components of the 
two Gaussian fit and left triangles, the narrow components.}
\label{dispersions}
\end{figure}

\section{Discussion}

\label{Velocity dispersions}

The upper panel of Fig.\ \ref{dispersions} shows the 
relation between the velocity dispersions of gas and stars. 
Circles correspond to gas velocity dispersions measured from the 
H$\beta$ emission line using a single Gaussian
fit. Squares and triangles correspond to measurements performed using
two-component Gaussian fits, squares for the broad  component and triangles
for the narrow one. The straight line shows the one-to-one relation. As a
general result, the H$\beta$ velocity dispersions derived by a single Gaussian
fit are lower than the stellar ones by about 25 km\,s$^{-1}$. This is also the
case for the Paschen lines in the two regions where they could be measured. On
the other hand, a good agreement is found between the stellar velocity
dispersions and those of the broad component of H$\beta$. The deviant point,
marked with arrows, corresponds to the region with the lowest signal-to-noise
ratio (R5) for which the fits do not provide accurate results. The narrow
component shows velocity dispersions even lower than those obtained by single
Gaussian fits. The ratio between the fluxes in the narrow and broad components
is between 0.7 and 0.95 (except for the case of R5 for which no meaningful
result is found). 

The lower panel of Fig.\ \ref{dispersions} shows the relation between the
stellar velocity dispersions and those of  the [O{\sc iii}] emission line
measured by both single and two-component Gaussian fits, which  due to the
weakness of the line (see figures 3 to 5), could be done in only four cases,
corresponding to the two slit positions on regions R2 and R3. 
In this case, the broad component seems to dominate the width of the emission
line which again agrees with the stellar one. The width of the narrow
component, in the cases in which the two-component fit was possible, is
comparable to that of the narrow component of the H$\beta$ line. 
 
\begin{figure}
\includegraphics[width=.48\textwidth,angle=0]{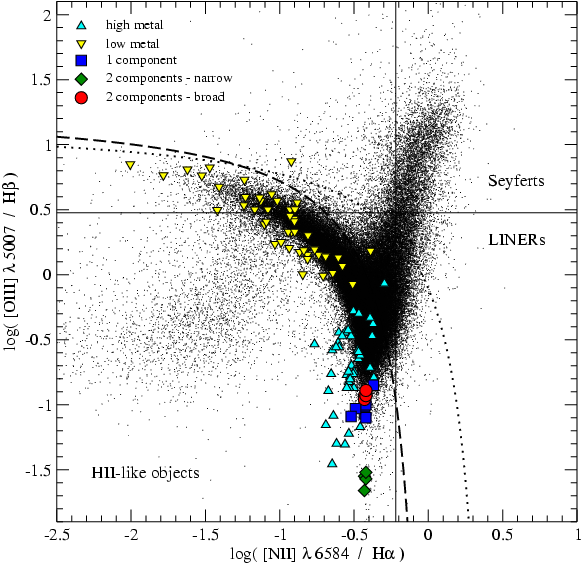}
\caption{BPT diagram [O{\sc iii}]/H$\beta$ {\it vs} [N{\sc
    ii}]/H$\alpha$. Squares correspond to the ratio of the emission 
  intensities of
  [O{\sc iii}] and H$\beta$ estimated using a single Gaussian fit, diamonds to
  the narrow components of the two Gaussian fits and circles to the broad
  components. Dotted and dashed curves are the boundary between
  Active Galactic Nuclei (AGNs) and \HII\ galaxies defined by Kewley et
  al. (2001) and Kauffmann et 
  al.\ (2003) respectively. The solid horizontal and vertical lines represent
  the division between Seyfert galaxies and LINERs according to Ho et
  al. (1997). Dots correspond to  a subsample of emission line objects,
  including HII galaxies, 
  from SDSS-DR3, from L\'opez (2005).  Triangles correspond to \HII\ regions
  from the sample of P\'erez-Montero \& D\'iaz (2005).
  They have been split into low metallicity (upside down triangles) and high
  metallicity regions (upward triangles) according to the criterion by D\'\i
  az \& P\'erez-Montero (2000) based on  oxygen and sulphur abundance
  parameters.}
\label{bpt}
\end{figure}

The two gaseous components have been plotted separately  in Fig.\ \ref{bpt}
which shows their location in the [O{\sc iii}]/H$\beta$ {\it vs} [N{\sc
    ii}]/H$\alpha$ diagram \cite*[BPT;][]{1981PASP...93....5B} together with a
sample of emission line 
galaxies (including \HII-like objects) from the SDSS-DR3 \citep{tesisjesus}
and \HII\ regions from the sample of 
\cite{2005MNRAS.361.1063P}.  The two systems are clearly segregated in
this diagnostic diagram with the narrow component showing the lowest
excitation of the two and occupying the same position in the diagram as the
starburst systems in the SDSS dataset with the lowest excitation found.

\begin{figure}
\centering
\includegraphics[width=.48\textwidth,angle=0]{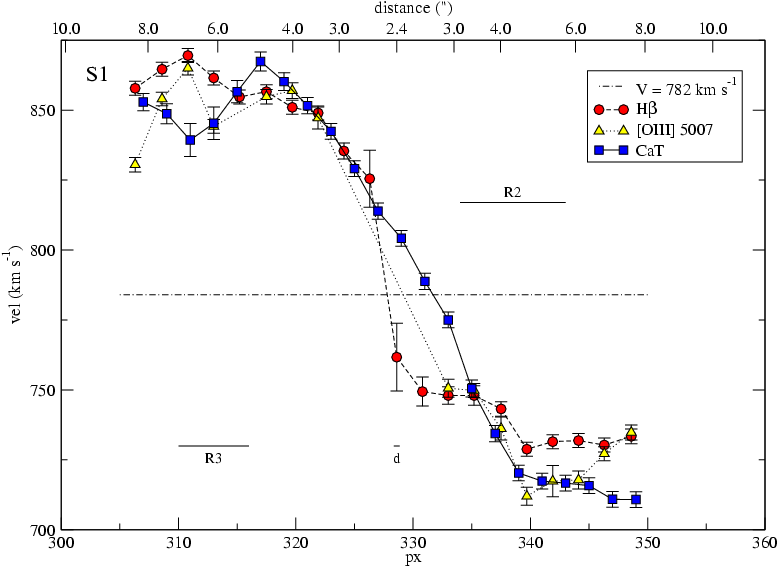}\vspace*{0.3cm}
\vspace*{0.3cm}
\includegraphics[width=.47\textwidth,angle=0]{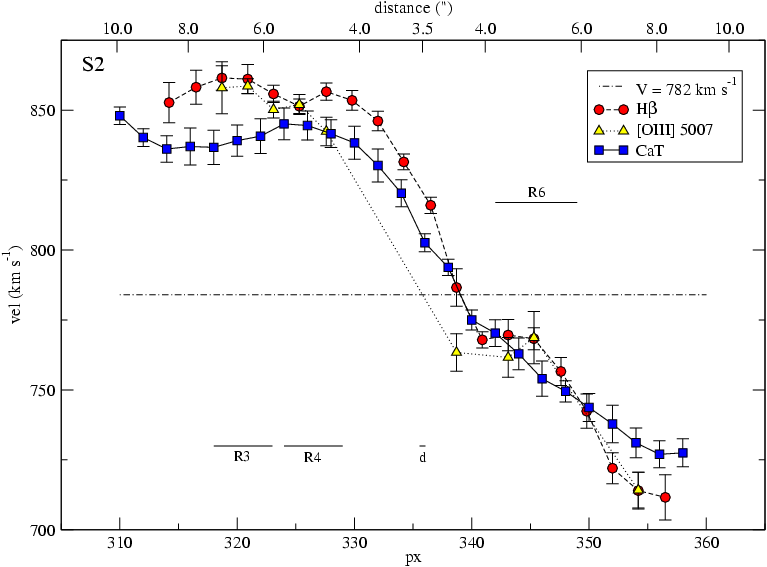}
\includegraphics[width=.48\textwidth,angle=0]{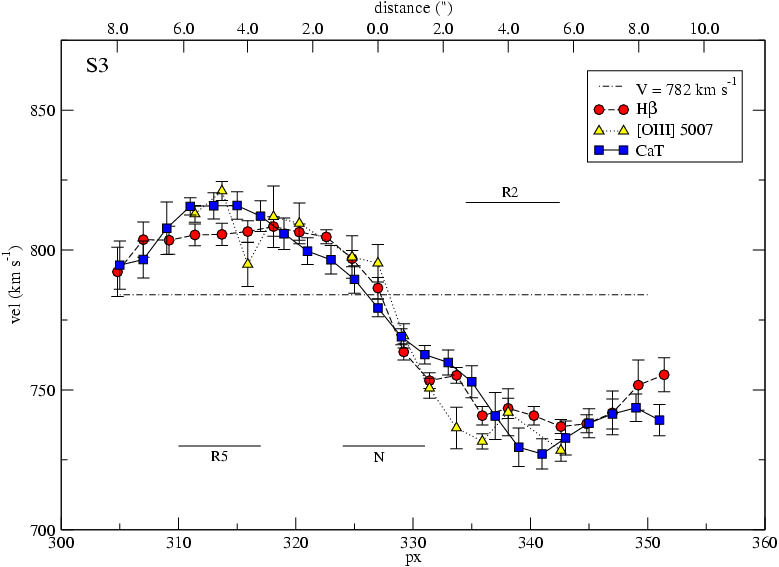}
\caption{Radial velocities  along the slit {\it vs} pixel number
  for each slit position (upper
  panel: S1; middle panel: S2; lower panel: S3) as derived
  from the gas emission lines (circles: H$\beta$; triangles: [O{\sc iii}])
  and the stellar absorption ones (squares). The individual CNSFRs and
  the nucleus, ``N'', or the closest position to it, ``d'', are marked in the
  plots. The dashed-dotted line is the systemic 
  velocity of NGC\,3351 derived by Planesas et al.\ (1997). The distance in
  arcsec from the nucleus is displayed in the upper x-axis of each panel.} 
\label{velocities}
\end{figure}

\label{Radial velocities}
Fig.\ \ref{velocities}  shows the radial velocities along the slit for
each slit position as derived from the ionised gas emission lines and stellar
absorptions. The turnover points of the rotational curves seem to be located 
at the same position than the starforming ring, specially for the S3 slit position that crosses the centre of the galaxy, as found in other galaxies
\citep[see][and references
  therein]{1988ApJ...334..573T,1999ApJ...512..623D}. For the systemic velocity
of NGC\,3351, the derived values are consistent with those previously obtained
by \cite{1975ApJ...199...39R} and Planesas et al.\ (1997), and with the
velocity  distribution expected for this type of galaxies
\citep{1987gady.book.....B}.  

The rotation velocities derived for both stars and gas are in reasonable
agreement, although in some cases the gas seems to rotate somewhat faster than
the stars. In fact, in the lower panel of Fig 13, which corresponds to the slit
position passing through the nucleus, it is interesting to note that at the
maximum and minimum of the velocity curve, which correspond approximately to
the positions of regions R5 and R2, the H$\beta$ emission line shows
velocities lower and higher than the stars by about 25 and 30\,km\,s$^{-1}$
respectively. This could be interpreted as motions of the ionised hydrogen
deviating from rotation and consistent with a radial infall to the central
regions of the galaxy. A similar result was found by Rubin et al.\ (1975) from
high dispersion observations of the H$\alpha$ line in the central region of
this galaxy. Their preferred model for the fitting of the kinematical data
consists of gas which is rotating and contracting with V$ _{rot}
$\,=\,126\,$\pm$\,16\,km\,s$^{-1}$ and V$ _{cont} $\,=\,34\,$ \pm
$\,11\,km\,s$^{-1}$. 

It  is clearly of major interest to find out how widespread is the presence of
two distinct components in the emission lines in ionised regions and what is
its influence on the observed line ratios 
for several reasons. Firstly, a change in position in the diagnostic diagrams
would certainly affect the classification of the activity in the central
regions of the concerned galaxies. Secondly,  it will affect the inferences
about the nature of the source of ionisation in the two components. And
thirdly, it could have an influence on the gas abundance determinations given
that the ratio of the narrow to the broad components of H$\beta$ is about
1. Clearly it is not possible to use global line ratios to estimate gaseous
abundances if the permitted and forbidden line fluxes are partially originated
in different kinematical systems. 
To disentangle the origin of these two components it will be necessary to map
these regions with high spectral and spatial resolution and much better S/N in
particular for the O$^{2+}$ lines.  3D spectroscopy with Integral Field Units
(IFUs) would be the ideal tool to approach this issue.  


Our values of the dynamical masses have been derived from stellar velocity
dispersion measurements as mapped by the CaT absorption lines and the sizes
measured on an HST image. They are in the range between 4.9\,$\times$\,10$^6$
and 4.3\,$\times$\,10$^7$\,M$_\odot$ for the CNSFRs and is
3.5\,$\times$\,10$^7$\,M$_\odot$ for the nuclear region inside the inner
11.3\,pc (see table \ref{mass}). The fractional errors of the dynamical masses
of the CNSFRs are between $\sim$3.4 and $\sim$\,17.8\,\%, and  is $\sim$3.1\,\%
for the nucleus. Masses derived from the H$\beta$ velocity dispersion under
the assumption of a single component for the gas would have been
underestimated  by factors between approximately 2 and 4. 
We have found that only two of the regions, R4 and R5, seem to possess
just one 
knot showing up in the continuum image and coincident with the H$\alpha$
emission. The sizes of these knots are 2.9 and 3.2\,pc respectively. The
rest of the regions are made up of several knots which presumably correspond
to individual star clusters. Their sizes are between 4.9 and 1.7\,pc, at the
limit of the resolution. For comparison, the size of cluster A in NGC\,1569 
is 1.9\,pc, as given by \cite{1995AJ....110.2665M}. The sizes and absolute
visual magnitudes estimated for each individual star cluster (e.g.
M$_v$\,=\,-12.55 for R4 derived from the HST image) are in the ranges 
established by \cite{1995AJ....110.2665M} in the definition of a super star
cluster (SSC). 

Other estimates of the masses of these regions have been obtained from the
fitting of broad band colours or spectra with the use of stellar population
synthesis models. Elmegreen et al.\ (1997) from near IR photometry in the J and
K bands and models by Leitherer \& Heckman (1995) for  instantaneous star
formation and solar and twice solar metallicity, derive masses of the
CNSFRs from 1 to 10\,$\times$\,10$^{5}$\,M$_\odot$. Colina
et al.\ (1997), from UV (IUE) spectra and instantaneous burst models by
\cite{1991A&AS...88..399M} derive a value of 3\,$\times$\,10$^5$\,M$_\odot$
for the whole SF ring. In both cases, in fact, these observables 
trace the young massive population which constitutes only part of the total 
mass. 

The masses derived here for the circumnuclear starforming individual clusters
of NGC\,3351 are between 1.8 and 8.7\,$\times$\,10$^6$\,M$_\odot$. These
values are 
between 5.5 and 26 times the mass derived for the SSC A in NGC\,1569 by
\cite{1996ApJ...466L..83H} from stellar velocity dispersion measurements using
red ($\sim$6000 \AA ) spectra, which is
(3.3$\pm$0.5)\,$\times$\,10$^5$\,M$_\odot$ and larger than
masses derived kinematically for SSC in irregular galaxies 
\citep*{2003ApJ...596..240M,2004AJ....128.2295L}.
They are also larger than those derived by \cite{2002AJ....123.1411B} for the
individual circumnuclear 
clusters of NGC\,4314 form HST imaging data following the procedure of
\cite{1999AJ....117..764E}, which are in the range 0.2\,$\times$\,10$^4 \leq$
M$\leq$\,1.6$\times$\,10$^4$\,M$_\odot$.  However, also the H$\beta$
luminosities of the NGC\,4314 clusters are lower than those of NGC\,3351 by a
factor of about 40.

The masses of the ionising stellar clusters of the CNSFRs, as derived from
their H$\alpha$ luminosities, are between 4.1\,$\times$\,10$^5$ and
2.4\,$\times$\,10$^6$\,M$_\odot$ for the starforming regions, and is
3.1\,$\times$\,10$^5$\,M$_\odot$ for the nucleus (see table
\ref{parameters}). These values are comparable to that derived by
Gonz\'alez-Delgado et al.\ (1994) for the circumnuclear region A in NGC\,7714 
(5.1$\times$10$^5$ M$_\odot$). In column 11 of table \ref{parameters} we show
a comparison (in percentage) between ionising stellar masses of the
circumnuclear regions  and their dynamical masses. These values vary
approximately in the range 1-11\,\% for the CNSFRs, and is 0.9\,\% for the
nucleus. 
Since the CaT absorption features are dominated by young stars (see
discussion above, Sec. \ref{masses}), and the M$_{ion}$/M$_{\ast}$ fraction is
still 
remarkably small in the case of the CNSFRs composed of single knots (R4 and
R5, for which the dynamical mass is most robustly estimated) we can assume 
that our upper limits to the dynamical masses, in spite of the limitations of
the method used to derive them, are rather tight. Then, our results concerning
the M$_{ion}$/M$_{\ast}$ fraction are robust.

Finally, the masses of the ionised gas vary between 3\,$\times$\,10$^3$ and
8.6\,$\times$\,10$^4$\,M$_\odot$ for the CNSFRs, and is
1\,$\times$\,10$^3$\,M$_\odot$ for the nucleus (see table \ref{parameters}),
also comparable to that derived by Gonz\'alez-Delgado et al. (1994) for the
circumnuclear region A in NGC\,7714 (3$\times$10$^5$ M$_\odot$). They make up
a small fraction of the total mass of the regions.  

Both the masses of the ionising stellar clusters and the ionised gas,  have
been derived from the H$\alpha$ luminosity of the CNSFRs assumed to consist of
one single component. However, if we consider only the broad component whose
kinematics follows that of the stars in the regions, all derived quantities
would be smaller by a factor of two.

\section{Summary and conclusions}

We have measured gas and stellar velocity dispersions in 5 CNSFRs and the
nucleus of the barred spiral  NGC\,3351. The stellar dispersions have been
measured from high resolution spectra of the CaT lines at
$\lambda\lambda$\,8494, 8542, 8662\,\AA, while the gas velocity dispersions
have been measured by Gaussian fits to the H$\beta$\,$\lambda$\,4861\,\AA\
line on high dispersion spectra. 

Stellar velocity dispersions are between 39 and 67\,km\,s$^{-1}$, about
20\,km\,s$^{-1}$ larger than those measured for the gas. However, the best
Gaussian fits involved two different components for the gas: a ``broad
component" with a velocity dispersion similar to that measured for the stars,
and a ``narrow component" with a dispersion lower than the stellar one by
about 30\,km\,s$^{-1}$. 

When plotted in a [O{\sc iii}]/H$\beta$ {\it vs} [N{\sc ii}]/H$\alpha$
diagram, the two systems are clearly segregated  with the narrow component
having the lowest excitation and being among the lowest excitation line ratios
detected within the SDSS dataset of starburst systems.

Values for the dynamical masses of the CNSFRs have been derived from stellar
velocity dispersions and are in the range between 4.9\,$\times$\,10$^6$ and
4.3\,$\times$\,10$^7$\,M$_\odot$ for the CNSFRs and is
3.5\,$\times$\,10$^7$\,M$_\odot$ for the nuclear region inside the inner
11.3\,pc. Masses derived from the H$\beta$ velocity dispersion under the
assumption of a single component for the gas would have been underestimated
by factors between approximately 2 to 4.

The derived masses for the individual clusters of NGC\,3351 are between 1.8
and 8.7\,$\times$\,10$^6$\,M$_\odot$. These values are between 5.5 and 26
times the mass derived for the SSC A in NGC\,1569 by
\cite{1996ApJ...466L..83H} and larger than other kinematically derived SSC
masses. 

Masses of the ionising stellar clusters of the CNSFRs have been derived from
their H$\alpha$ luminosities under the assumption that the regions are
ionisation bound and without taking into account any photon absorption by
dust. Their values are between 4.1\,$\times$\,10$^5$ and
2.4\,$\times$\,10$^6$\,M$_\odot$ for the starforming regions, and is
3.1\,$\times$\,10$^5$\,M$_\odot$ for the nucleus (see table \ref{parameters}),
comparable to that derived by Gonz\'alez-Delgado et al.\ (1994) for the
circumnuclear region A in NGC\,7714. Therefore, the ratio of the ionising
stellar population to the total dynamical mass is between 0.01 and 0.11. 

Derived masses for the ionised gas vary between 3\,$\times$\,10$^3$ and
8.6\,$\times$\,10$^4$\,M$_\odot$ for the CNSFRs, and is
1\,$\times$\,10$^3$\,M$_\odot$ for the nucleus, also comparable to that
derived by Gonz\'alez-Delgado et al.\ (1994).

It is interesting to note that, according to our findings, the SSC in CNSFRs
seem to contain composite stellar populations. Although the youngest one
dominates the UV light and is responsible for the gas ionisation, it
constitutes only about 10\,\% of the total. This can explain the low
equivalent 
widths of emission lines measured in these regions.  This may well apply to
the case of other SSC and therefore conclusions drawn from fits of SSP (single
stellar population) models should be taken with caution 
\citep*[e.g.][]{2003ApJ...596..240M,2004AJ....128.2295L}.
Also the composite
nature of the CNSFRs  means that star formation in the rings is a process that
has taken place over time periods much longer than those implied by the
properties of the ionised gas.

The rotation velocities derived for both stars and gas are in reasonable
agreement, although in some cases the gas shows a velocity slightly different
from that of the stars. The rotation curve corresponding to the position
going through the centre of the galaxy shows maximum and minimum values at the
position of the circumnuclear ring, as observed in other galaxies with
CNSFRs. The differences in velocity between gas and stars can be interpreted
as motions of the ionised hydrogen deviating from rotation and consistent with
a radial infall to the central regions of the galaxy. Our results are
consistent with those found by Rubin et al.\ (1975) and would yield an infall
velocity of about 25\,km\,s$^{-1}$.\\ 

The existence of more than one velocity component in the ionised
gas corresponding to kinematically distinct systems, deserves further
study. Several results derived from the observations of the different emission
lines could be affected, among others: the classification of the activity in
the central regions of galaxies, the inferences about the nature of the source
of ionisation, the gas abundance determinations, the number of ionising
photons from a given region and any quantity derived from them, etc. To
disentangle the origin of these two components it will be necessary to map
these regions with high spectral and spatial resolution and much better S/N in
particular for the O$^{2+}$ lines.  3D spectroscopy with Integral Field Units
(IFUs) would be the ideal tool to approach this issue. \\

{\it Acknowledgments} 
\label{Acknoledgement}
We are indebted to Jes\'us L\'opez who provided the data on the SDSS sample
prior to publication. We acknowledge fruitful discussions with Horacio
Dottori, Enrique P\'erez, Enrique P\'erez-Montero and Jos\'e V\'ilchez.   
We thank very much an anonymous referee for his/her careful examination
of our manuscript. We have found the report extremely thorough and undoubtedly
has helped to improve the contents of this paper.

The WHT is operated in the island of La Palma by the Isaac Newton Group
in the Spanish Observatorio del Roque de los Muchachos of the Instituto
de Astrof\'\i sica de Canarias. We thank the Spanish allocation committee
(CAT) for awarding observing time.

Some of the data presented in this paper were obtained from the
Multimission Archive at the Space Telescope Science Institute (MAST). STScI is
operated by the Association of Universities for Research in Astronomy, Inc.,
under NASA contract NAS5-26555. Support for MAST for non-HST data is provided
by the NASA Office of Space Science via grant NAG5-7584 and by other grants
and contracts.

This research has made use of the NASA/IPAC Extragalactic Database (NED) which
is operated by the Jet Propulsion Laboratory, California Institute of
Technology, under contract with the National Aeronautics and Space
Administration. 

This research has made use of the SIMBAD database, operated at CDS,
Strasbourg, France. 

This work has been supported by DGICYT grant AYA-2004-02860-C03. GH and MC acknowledge support from the Spanish MEC through FPU grants AP2003-1821 and AP2004-0977. AID acknowledges support from  the Spanish MEC through a sabbatical grant PR2006-0049. Also, partial support from the Comunidad de Madrid under grant S-0505/ESP/000237 (ASTROCAM) is acknowledged.Support from the Mexican Research Council (CONACYT) through grant 49942 is acknowledged by ET and RT. We thank the hospitality of the Institute of Astronomy of Cambridge where this paper was written.

\bibliographystyle{mn2e}
\bibliography{ngc3351}

\end{document}